\documentclass[a4paper,11pt]{article}
\pdfoutput=1 

\usepackage{jheppub} 

\usepackage{amsmath}
\usepackage{amsfonts,amssymb}

\usepackage[T1]{fontenc} 
\usepackage{bm}

\def\({\left(} 
\def\){\right)}

\usepackage{xcolor}
\xdefinecolor{redviolet}{RGB}{228,38,207}
\xdefinecolor{brickred}{RGB}{220,60,66}
\xdefinecolor{bluscuro}{RGB}{51,51,179}

\usepackage{hyperref}

\title{\boldmath Muon--electron scattering at NLO}

\author[a,b]{Massimo Alacevich,}
\author[b]{Carlo M. Carloni Calame,}
\author[c]{Mauro Chiesa,}
\author[a,b]{Guido Montagna,}
\author[b]{Oreste Nicrosini}
\author[b]{and Fulvio Piccinini}

\affiliation[a]{Dipartimento di Fisica, Universit\`a di Pavia, Via A. Bassi 6, 27100 Pavia, Italy}
\affiliation[b]{INFN, Sezione di Pavia, Via A. Bassi 6, 27100 Pavia, Italy}
\affiliation[c]{Institut f\"ur Theoretische Physik und Astrophysik, Julius-Maximilians-Universit\"at W\"urzburg,\\
Emil-Hilb-Weg 22, D-97074 W\"urzburg, Germany}

\emailAdd{massimo.alacevich@pv.infn.it}
\emailAdd{carlo.carloni.calame@pv.infn.it}
\emailAdd{mauro.chiesa@physik.uni-wuerzburg.de}
\emailAdd{guido.montagna@pv.infn.it}
\emailAdd{oreste.nicrosini@pv.infn.it}
\emailAdd{fulvio.piccinini@pv.infn.it}

\abstract{We consider the process of muon-electron elastic scattering,
  which has been
  proposed as an ideal framework to measure the running of the electromagnetic coupling constant 
at space-like momenta and determine the leading-order hadronic contribution to 
the muon $g-2$ (MUonE experiment). 
We compute the next-to-leading (NLO) contributions due to QED and purely weak 
corrections and implement them into a fully differential Monte Carlo
event generator, which is
available for first experimental studies. We show representative phenomenological 
results of interest for the MUonE experiment and examine in detail the impact of the
various sources of radiative corrections under different selection criteria, 
in order to study the dependence of the NLO contributions on the
applied cuts. The study represents the first step towards
the realisation of a high-precision Monte Carlo code necessary for data analysis.}

\keywords{Fixed target experiments, Precision QED, NLO computations}

\begin{document} 
\maketitle
\flushbottom

\section{Introduction}
\label{sec:intro}

The anomalous magnetic moment of the muon $a_\mu = (g - 2)_\mu / 2$
represents a formidable test of the
Standard Model (SM) of particle
physics~\cite{Jegerlehner:2017gek,Jegerlehner:2009ry}.
There is a persisting tension between the measured value, as provided
by the BNL-E821 experiment with an accuracy of
0.54ppm~\cite{Bennett:2006fi}, and the SM prediction, which presently
exceeds the $3\sigma$
level~\cite{Jegerlehner:2018zrj,Knecht:2018nhq,Keshavarzi:2018mgv,Davier:2016udg}. 
This discrepancy could be ascribed to some uncontrolled experimental
or theoretical systematics or could signal the presence of New Physics
beyond the SM. Two next-generation muon $g-2$ experiments at
Fermilab~\cite{Grange:2015fou} and J-PARC~\cite{Mibe:2011zz} plan to improve the 
experimental error by a factor of four, which constitutes a challenge
for the theory and has triggered a lot of new theoretical
efforts~\cite{Jegerlehner:2018zrj,Knecht:2018nhq}.

The calculation of $a_\mu$ in the SM requires the evaluation of subtle quantum loop effects 
and its present uncertainty, at the level of $3.5 \times 10^{-10}$, is dominated by strong interaction
contributions, which can not be computed perturbatively at low
energies. The main sources of error
come from the evaluation of the leading-order hadronic correction, $a_\mu^\text{HLO}$, and 
of the hadronic light-by-light contribution. In particular, the most precise 
estimates of $a_\mu^\text{HLO}$ rely on the calculation of a dispersion integral of the 
hadron production cross section measured in $e^+ e^-$ annihilation at a few GeV scale.
This data-driven determination of the $a_\mu^\text{HLO}$ contribution to the $g-2$ is 
affected by uncertainties coming from the experimental error, as well
as from delicate issues in the treatment and combination of different
data corresponding to many exclusive 
channels~\cite{Jegerlehner:2018zrj,Keshavarzi:2018mgv,Davier:2016udg}. Hence,
alternative and independent approaches to the standard evaluation of
$a_\mu^\text{HLO}$ via time-like data are of utmost importance to shed
light on the situation.\footnote{Recent progress in lattice QCD calculations 
of $a_\mu^\text{HLO}$ is documented in
Refs.~\cite{DellaMorte:2017dyu,Wittig:2018r,Borsanyi:2017zdw,Blum:2018mom,Giusti:2018mdh}.}

In this respect, it has been proposed in recent times~\cite{Calame:2015fva} to derive the leading-order hadronic contribution 
to the muon anomaly from a measurement
of the effective electromagnetic coupling in the 
space-like region (and in particular of its hadronic contribution $\Delta\alpha_\text{had}(q^2)$),
following ideas first put forward in Ref.~\cite{Lautrup:1971jf}. Shortly afterwards, the 
process of muon-electron elastic scattering, i.e. $\mu e \to \mu e$, as measurable by 
scattering high-energy muons on electrons at rest, has been recognised 
as an ideal process to perform such 
a measurement~\cite{Abbiendi:2016xup}.\footnote{A method to measure the 
running of the QED coupling in the space-like region using small-angle 
Bhabha scattering was proposed in Ref.~\cite{Arbuzov:2004wp} and applied 
to LEP data by OPAL Collaboration~\cite{Abbiendi:2005rx}.}
 An experiment devoted to the 
measurement of $\alpha_\text{QED} (t)$, where $t$ is the space-like squared momentum transfer,
via $\mu e$ scattering (MUonE) is presently under consideration at CERN 
in the context of the {\it Physics Beyond Colliders} 
study group.\footnote{See
  \href{https://indico.cern.ch/category/7939}{\tt https://indico.cern.ch/event/644287/}} The idea 
consists in performing a fixed-target experiment, by using a 150~GeV high-intensity muon beam, 
presently available at the CERN North area, incident on atomic electrons of a low-$Z$ target.

The challenge of the proposed experiment is the feasibility of achieving a statistical and
systematic uncertainty in the measurement of the $\mu e$ differential cross sections 
at the level of 10ppm. If this task is accomplished, the new space-like determination 
of the leading hadronic contribution to the muon $g-2$ will turn out to be competitive with 
the present time-like approach. In addition to the systematics of experimental 
nature, the calculation of the radiative corrections to the $\mu e$ cross section represents
a further source of systematic uncertainty to be taken carefully under control. Ultimately, this calculation
will be necessary according to the highest standards of theoretical precision, in order to 
cope with the very demanding requirement of accuracy.

In this paper, we take a first step towards the calculation of higher-order contributions to
$\mu e \to \mu e$ scattering, by computing the full set of NLO corrections to the process in the SM as
due to QED and purely weak effects. 
We calculate the above corrections without any approximation, i.e. retaining finite lepton 
(muon and electron) mass
contributions, and make them available in the form of a fully flexible
Monte Carlo (MC) code, 
which is unavoidable to analyse the precision data of the MUonE experiment. We also scrutinise 
the size of the various sources of radiative corrections, in order to provide a guideline to future
efforts in the calculation of NNLO corrections and resummation.

The NLO QED corrections to the $\mu e$ scattering cross section were computed 
long time
ago~\cite{Nikishov:1961,Eriksson:1961,Erikssonetal:1964,VanNieuwenhuizen:1971yn,DAmbrosio:1984abj,Kukhto:1987uj,Bardin:1997nc}
and recently revisited in Ref.~\cite{Kaiser:2010zz}. However, the
above calculations contain some assumptions (vanishing 
electron mass, soft photon approximation), which do not apply to the present work.
To the best of our knowledge, one-loop weak corrections,
due to the presence of electroweak bosons in the internal loops, have
never been considered so far. Just the contribution of the $\gamma$-$Z$
interference at the tree-level was computed in
Ref.~\cite{DAmbrosio:1984abj} and compared to $O(\alpha)$ QED
corrections at relatively high energies. 

{On the NNLO side}, the master integrals for the two-loop planar and non-planar box
diagrams in QED have been evaluated in Ref.~\cite{Mastrolia:2017pfy}
and Ref.~\cite{DiVita:2018nnh}, respectively, retaining full
dependence on the muon mass. NNLO hadronic corrections to $\mu e$
scattering, given by diagrams with a hadronic vacuum polarisation
insertion in the photon propagator, have been computed recently in Ref.~\cite{Fael:2018dmz,Fael:2019nsf}.
Further theoretical progress in 
the calculation of NNLO corrections and resummation was presented at the 
MITP topical workshop {\it ``The evaluation of the leading hadronic contribution to the 
muon anomalous magnetic moment''}, held in Mainz in February 2018.\footnote{See the 
material available at the link \href{https://indico.mitp.uni-mainz.de/event/128/}{\tt https://indico.mitp.uni-mainz.de/event/128/}}

The paper is organised as follows. In Section~\ref{sec:theory} we describe the details of the NLO calculation.
In Section~\ref{sec:numerics} we present and discuss our phenomenological results, both 
at the level of integrated cross sections (Section~\ref{sec:integrated}) and 
differential observables (Section~\ref{sec:differential}). We study in particular the impact of the NLO
corrections as a function of the imposed event selection criteria and scrutinise the r\^ole played
by the different gauge-invariant subsets of corrections. In Section~\ref{sec:conclusions} we draw our conclusions and 
discuss the perspectives of our work.

\section{Details of the NLO calculation}
\label{sec:theory}

In the present Section, we describe the details of our theoretical
approach. We first discuss the computation of the NLO QED and
electroweak contributions without any approximation, i.e. retaining full dependence 
on the lepton masses. Next, we mention the procedure followed for the calculation of the QED 
corrections in the limit of vanishing electron mass, which is sketched
in Section~\ref{sec:numerics}.

\subsection{NLO QED corrections}
\label{sec:qedexact}

The differential {unpolarised} cross section of muon-electron
elastic scattering
\begin{eqnarray*}
\mu^{\pm} (p_1) + e^- (p_2) \to \mu^{\pm} (p_3) + e^- (p_4) 
\end{eqnarray*}
at LO in QED is given by 
\begin{equation}
\frac{d\sigma}{d t} \, = \, \frac{1}{\lambda(s, m_\mu^2, m_e^2)} \, \frac{4\pi \alpha^2}{t^2} \left[ (s - m_\mu^2 - m_e^2)^2  + st + \frac{1}{2} t^2\right]
\label{eq:xslo}
\end{equation}
In Eq.~(\ref{eq:xslo}) $\alpha$ is the fine structure constant,
 $\lambda$ is the K\"all\'en function and $s$, $t$ are the usual Mandelstam variables, 
which are {calculated with massive four momenta of the incoming
  and outgoing muon ($p_1$ and $p_3$) and of the incoming and outgoing
  electron ($p_2$ and $p_4$).}
In our calculation, we consider the scattering initiated by both positive and negative muons, both options being possible in the 
muon beam available at the CERN North Area.

In the electroweak theory, the NLO corrections to the process consist of QED and purely weak contributions. Since $\mu e$ scattering is 
a neutral-current process, the two subsets are separately gauge invariant and can be treated separately. A priori, the QED corrections 
are expected to be the dominant contribution, as photons which are emitted collinear to a lepton give rise to 
enhanced logarithmic 
corrections of the form $\alpha \log (Q^2 / m_\ell^2)$ 
in the large $Q^2$ limit, where $m_\ell$ is the lepton mass and $Q$ some typical energy scale.

We calculate the $2\to 2$ amplitude including one-loop virtual
  corrections in the on-shell renormalization scheme,
whereas real photon corrections are induced by the $2\to 3$ bremsstrahlung  
process $\mu e \to \mu e + \gamma$. The latter contribution is infrared (IR) divergent, but its sum with vertex and 
box corrections is IR-finite. We regularise the IR singularities according to the standard QED procedure of assigning a 
vanishingly small mass to the photon in the computation of the virtual and real contributions. Ultraviolet (UV)
divergences associated to loop diagrams are treated using dimensional regularisation.

In formulae, the  NLO cross section is split into two contributions and computed as follows
\begin{equation}
\sigma_\text{NLO} \, = \,  \sigma_{\mu e \to \mu e} + \sigma_{\mu e \to \mu e \gamma} \, \equiv \, \sigma_{2 \to 2} + \sigma_{2 \to 3}
\label{eq:xsnlo}
\end{equation}
In Eq.~(\ref{eq:xsnlo}), $\sigma_{2 \to 2}$ is given by the sum of the LO cross section and the NLO one containing virtual photonic corrections and 
reads explicitly as follows
\begin{equation}
\sigma_{2 \to 2} \, = \, \sigma_\text{LO} + \sigma_\text{NLO}^\text{virtual} = \frac{1}{F} \int d \Phi_2 \left( |{\cal{A}}_\text{LO}|^2 + 
2 \, \text{Re} [{\cal{A}}_\text{LO}^\dagger \times {\cal{A}}_\text{NLO}^\text{virtual} (\lambda)] \right)
\end{equation}
where
$F=2\sqrt{(s-m_e^2-m_\mu^2)^2-4m_e^2m_\mu^2}$
is the incoming flux factor, $d \Phi_2$ the $2 \to 2$ phase space
volume and $\lambda$ the fictitious photon mass.

For the calculation of the $2 \to 3$ real photon contribution, we
introduce a phase space slicing in terms of an arbitrarily small cutoff on 
the photon energy $\omega$ and compute $\sigma_{2 \to 3}$ according to the following formula
\begin{equation}
\sigma_{2 \to 3} \, = \, \frac{1}{F} \int_{\omega > \lambda} d \Phi_3 \, | {\cal{A}}_\text{NLO}^{1\gamma} |^2 =
\frac{1}{F} \left( \int_{\lambda < \omega < \omega_s} d \Phi_3 \, |
     {\cal{A}}_\text{NLO}^{1\gamma} |^2 + \int_{\omega > \omega_s} d \Phi_3 
\, | {\cal{A}}_\text{NLO}^{1\gamma} |^2 \right)
\label{eq:xsnlor}
\end{equation}
where $\omega_s$ is a soft-hard slicing separator, with $\lambda\ll\omega_s\ll\sqrt{s}$, and 
$d \Phi_3$ is the $2 \to 3$ phase space element. The first
  contribution in the r.h.s of Eq.~(\ref{eq:xsnlor}) is calculated retaining the photon mass $\lambda$,
  while in the second one the photon is treated as massless. By integrating the
first contribution analytically, Eq.~(\ref{eq:xsnlor}) 
can be recast as follows
\begin{equation}
\sigma_{2 \to 3} \, = \, \Delta_s (\lambda, \omega_s) \int d \sigma_\text{LO} + \frac{1}{F} \int_{\omega > \omega_s} d \Phi_3 
\, | {\cal{A}}_\text{NLO}^{1\gamma} |^2 
\end{equation}
where $\Delta_s (\lambda, \omega_s)$ is the eikonal factor for the real radiation correction in the soft photon approximation, whose 
expression can be found for instance in Ref.~\cite{Bohm:2001yx}. 

The contribution due to vacuum polarisation is taken into account using 
the effective charge as an overall factor through the replacement $\alpha \to \alpha(t)$ in the LO and NLO cross section. 
The running of the QED coupling is calculated using the following 
Dyson-resummation expression
\begin{equation}
\alpha (t) \, = \, \frac{\alpha}{1 - \Delta\alpha(t)}
\label{eq:alpharun}
\end{equation}
where
\begin{equation}
\Delta\alpha (t) \, = \, \Delta\alpha_\text{lep} (t) +  \Delta\alpha_\text{top} (t) + \Delta\alpha_\text{had} (t)
\label{eq:dalpha}
\end{equation}
In this work, we implement the leptonic correction
$\Delta\alpha_\text{lep}$ and
the (tiny) top-quark contribution $\Delta\alpha_\text{top}$ in
one-loop approximation\footnote{Note however that the leptonic contribution is
  known at three \cite{Steinhauser:1998rq} and at four
  loops~\cite{Sturm:2013uka} {in QED}:
  the implementation in the present calculation of higher-order
  corrections is straightforward and their impact is studied in
  Section~\ref{sec:differential}. {We also refrain from including any
  weak correction beyond one-loop approximation (see e.g.~\cite{Degrassi:2003rw}),
  because they are found to be negligible already at NLO, as
  discussed in Section~\ref{sec:differential}.}} and the hadronic 
correction is taken into account according to the dispersive approach implemented in the latest version 
of the {{\tt hadr5n16.f} routine}~\cite{Jegerlehner:2017zsb}.\footnote{Available at \href{http://www-com.physik.hu-berlin.de/~fjeger/software.html}{\tt http://www-com.physik.hu-berlin.de/\~{}fjeger/software.html}}
Note, however, that we use the value $\alpha = \alpha (0)$ of the fine
structure constant for the coupling of the photon 
to the external charged legs. {We would like also to emphasize
  that we use here the {\tt hadr5n16.f} routine only to get the expected
  effect on the running of $\alpha$ induced by
  $\Delta\alpha_\text{had} (t)$, which will be extracted from MUonE
  data by inserting Eq.~(\ref{eq:dalpha}) into Eq.~(\ref{eq:alpharun}) and
solving for $\Delta\alpha_\text{had} (t)$.}

Concerning the method of calculation and related cross-checks, all the Feynman diagrams 
for virtual and real photon contributions were manipulated with the 
help of the symbolic manipulation program \textsc{Form}~\cite{Ruijl:2017dtg,Kuipers:2012rf}, retaining full dependence on fermion masses and helicities. 
The evaluation of one-loop tensor coefficients and scalar 2/3/4 points functions was performed 
using the package \textsc{LoopTools}~\cite{Hahn:2010zi,Hahn:1998yk} and cross-checked against the 
\textsc{Collier} software~\cite{Denner:2016kdg}, 
finding perfect agreement between the results of the two libraries.
{UV-finiteness and independence of the $\lambda$ parameter have
  been checked analytically while the independence on $\omega_s$
  has been verified with high numerical accuracy in the range
  $10^{-7}$~MeV~$< \omega_s
  < 10^{-3}$~MeV.}

\subsection{NLO electroweak corrections}
\label{sec:ewexact}
The calculation of the full set of NLO corrections in the electroweak theory has been performed according to the 
following details: the tree-level and one-loop matrix elements for the virtual and real corrections have 
been evaluated without any approximation with the computer program \textsc{Recola}~\cite{Actis:2016mpe}, 
which internally uses the \textsc{Collier}
library for the one-loop scalar~\cite{tHooft:1978jhc,Beenakker:1988jr,Dittmaier:2003bc,Denner:2010tr} 
and tensor integrals~\cite{Passarino:1978jh,Denner:2002ii,Denner:2005nn}. In accordance with the 
calculation of the QED corrections, soft-photon singularities have been regularised in terms of an infinitesimal 
photon mass parameter and renormalization has been performed in the on-shell scheme. 
In order to compare consistently with the results of the QED calculation at the level of purely photonic corrections, 
the fermionic one-loop contributions to the photon propagator, which
are automatically computed by the \textsc{Recola} program, have been calculated analytically and
subtracted from the \textsc{Recola} output.

\subsection{QED corrections in the vanishing electron mass limit}
\label{sec:mezero}

The results for the NLO corrections discussed above  
include the contributions due to finite lepton mass effects, such as corrections of the
kind $\alpha \, m_\ell^2 / Q^2 \, \log (Q^2 / m_\ell^2)$ and $\alpha \, m_\ell^2 / Q^2$. 
As a consequence of the available c.m. energy and typical values of the squared 
momentum transfer relevant for the MUonE experiment, muon mass effects can not be neglected and contribute, 
together with logarithmic enhanced terms, to make the whole NLO muon correction.
However, also because of the complexity of a two-loop calculation in the presence of 
two different massive
fermions~\cite{Mastrolia:2017pfy,DiVita:2018nnh,Engel:2018fsb}, it is worth
assessing the r\^ole played by the contributions induced by a finite
electron mass within the one-loop calculation.

To this end, we performed a calculation of the NLO QED corrections by treating the
electron as massless, wherever possible. The adopted approximation is detailed
in Section~\ref{sec:numerics}.

\section{Numerical results}
\label{sec:numerics}

In this Section, we show and discuss the phenomenological results obtained by using a 
fully differential MC code which implements the theoretical
approach described in Section~\ref{sec:theory}. For definiteness, we
remark that all calculations and simulations are performed in the
center of mass (c.m.) frame and then the momenta are boosted to the
laboratory frame, where the initial-state electron is at rest.

We present numerical results both for the $\mu^- e^- \to \mu^- e^-$
and $\mu^+ e^- \to \mu^+ e^-$ process, since both options are relevant for the MUonE experiment and the radiative corrections differ in the two cases, as it will be shown in the following.

For the QED calculation, the input parameters are set to:
\begin{eqnarray}
  &\alpha(0)\equiv\alpha = 1/137.03599907430637&\nonumber\\
  &m_e = 0.510998928\text{ MeV} \qquad m_\mu = 105.6583715\text{ MeV}&\nonumber\\
  &m_\tau = 1.77682\text{ GeV}\qquad m_\text{top} = 175.6\text{ GeV}&
\label{eq:inputqedparams}
\end{eqnarray}

{The electroweak predictions are obtained in a scheme where $\alpha$, $M_W$ and $M_Z$
are input parameters. Besides the inputs of
Eq.~(\ref{eq:inputqedparams}), NLO electroweak predictions provided by
\textsc{Recola} further require
$M_Z=91.1876\text{ GeV}$, $M_W=80.376\text{ GeV}$,
$M_H=125\text{ GeV}$, $\Gamma_Z=\Gamma_W=0\text{ GeV}$ and the masses
of the light quarks. The latter are needed for the calculation of the
fermionic one-loop corrections to the photon and $Z$ propagators. As
remarked above, the fermionic one-loop corrections to the photon propagator are
analytically subtracted, leaving a tiny dependence of the full NLO
electroweak result on the light quark masses {in the $Z$ self energy and $\gamma/Z$
mixing (which is anyway not singular as the quark masses go to zero)}. Their value is set
in such a way to reproduce $\Delta\alpha_\text{had}^{(5)}(M_Z^2)$ (see
for instance~\cite{Alioli:2016fum}).}

For the energy of the incoming muons, we assume $E_\mu^\text{beam} = 150$~GeV, which is the 
energy of the M2 beam line of the CERN SPS. Note that, under the
fixed-target configuration of the MUonE experiment, the 
c.m. energy corresponding to this muon energy is given by
$\sqrt{s}\simeq 0.405541$~GeV {and that the Lorentz $\gamma$ factor
  boosting the c.m. frame into the lab frame is $\gamma\simeq 370$}.
{We stress that in the lab frame
the relation $t_{ee}\equiv (p_2-p_4)^2 = 2m_e^2 - 2m_eE_e$
(where $E_e$ is the energy of the scattered electron) holds, which
implies that a lower limit on $E_e$ corresponds to an upper limit
on $t_{ee}$.}

In this kinematical condition, the collinear logarithms $L_e = \ln (s/m_e^2)$ and 
$L_\mu = \ln (s/m_\mu^2)$ are of the order of $L_e \simeq 13.4$ and $L_\mu \simeq 2.7$, respectively.

In order to study the dependence of the radiative corrections on the applied cuts, we consider 
four different event selections defined by the following criteria:
\begin{enumerate}

\item  $\theta_e,\theta_\mu<100\text{\ mrad}$ and $E_e > 0.2$~GeV
  {(i.e. $t_{ee}\lesssim -2.04\cdot 10^{-4}\text{ GeV}^2$)}. The angular cuts model the typical acceptance 
conditions of the experiment and the electron energy threshold is imposed to guarantee the presence of two
charged tracks in the detector;

\item $\theta_e,\theta_\mu<100\text{\ mrad}$ and $E_e > 1$~GeV {
  (i.e. $t_{ee}\lesssim -1.02\cdot 10^{-3}\text{ GeV}^2$)}. With respect to Setup 1, a higher 
  electron energy threshold is imposed to
  {focus on the region where
  $\Delta\alpha_\text{had}(t)$ is larger};

\item the same criteria as in Setup 1, with an additional acoplanarity cut, 
applied to partially remove radiative events and thus enhancing the
fraction of elastic events. We require acoplanarity~$\equiv|\pi -
(\phi_e-\phi_\mu)|\le 3.5\text{\ mrad}$ for the sake of illustration;

\item the same criteria as in Setup 2, with the additional acoplanarity cut 
$|\pi - (\phi_e-\phi_\mu)|\le 3.5\text{\ mrad}$. 
\end{enumerate}
\begin{figure}[ht]
\begin{center}
\includegraphics[width=0.7\textwidth]{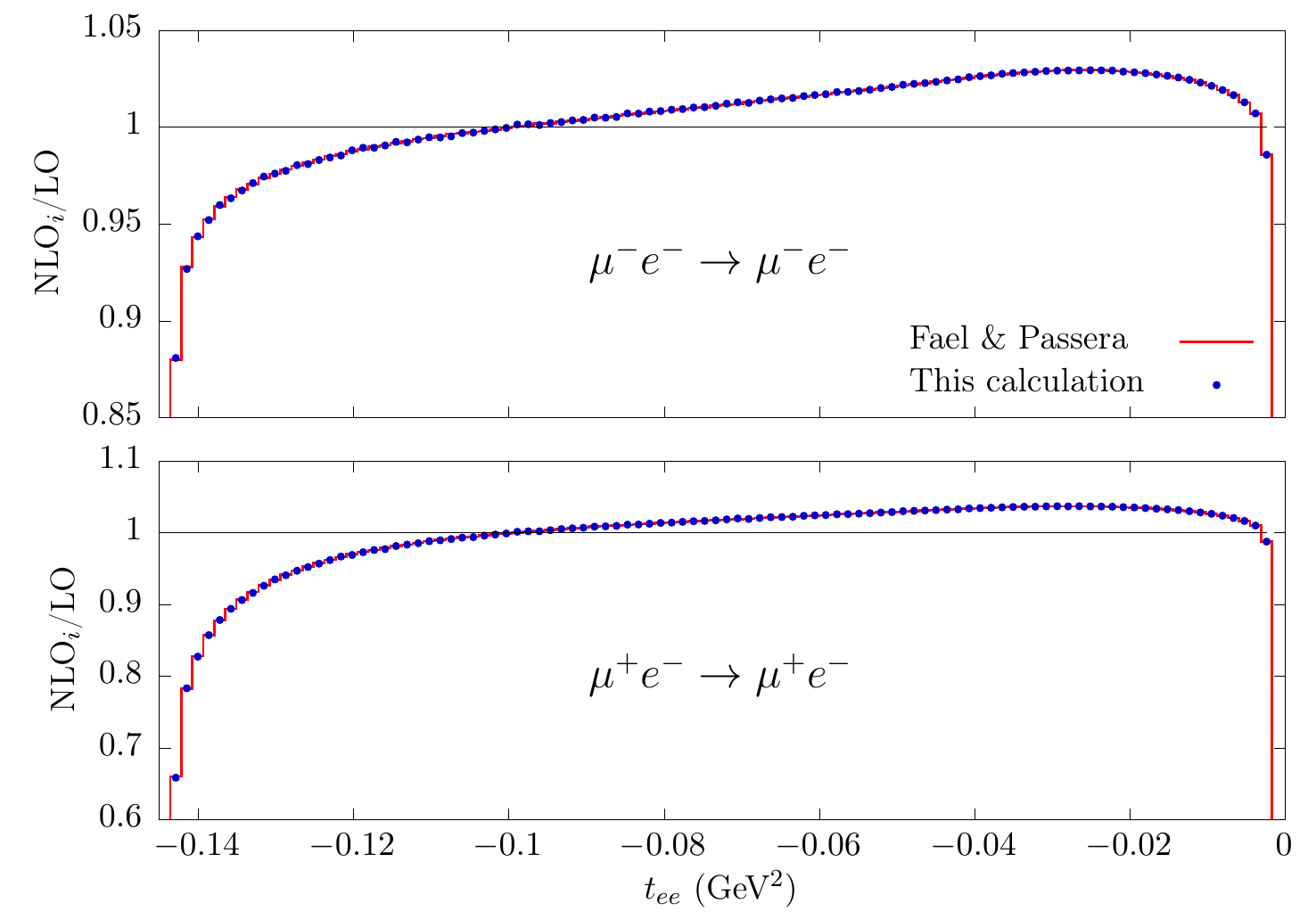}
\end{center}
\caption{\label{Fig:Fig1} Tuned comparison between our predictions and those of Ref.~\cite{Fael:2018pc} 
for the ratio of the NLO and LO cross section in QED
of the $\mu^- e^- \to \mu^- e^-$ (upper panel) and $\mu^+ e^- \to \mu^+ e^-$ (lower panel) process, as
a function  of the squared momentum transfer $t_{ee} = (p_2 - p_4)^2$.}
\end{figure}
Notice that $\Delta\alpha_\text{had}(t)$ increases as $|t|$
increases and thus, in this kinematical configuration, its largest
value corresponds to the small $\theta_e$ region.
It is also worth noting that all the numerical results are given for so-called bare
leptons, i.e. in the absence of lepton-photon recombination criteria,
as the details of the measurement of the lepton energies are still
under study by the MUonE experiment.

Before entering the discussion of our phenomenological analysis,
we show in Fig.~\ref{Fig:Fig1} the results of a tuned comparison 
(i.e. using the same set of input parameters and cuts) between our
predictions and those of Ref.~\cite{Fael:2018pc}. The comparison has
been performed in order to test the technical accuracy of 
our calculation and its MC implementation. The results shown
in Fig.~\ref{Fig:Fig1} correspond {to the ratio of the NLO QED differential cross 
section and the LO one, both for} the process $\mu^- e^- \to \mu^- e^-$
(upper panel) and $\mu^+ e^- \to \mu^+ e^-$ (lower panel), as a
function of the squared momentum transfer $t_{ee}$. 

\begin{figure}[ht]
\begin{center}
\includegraphics[width=0.7\textwidth]{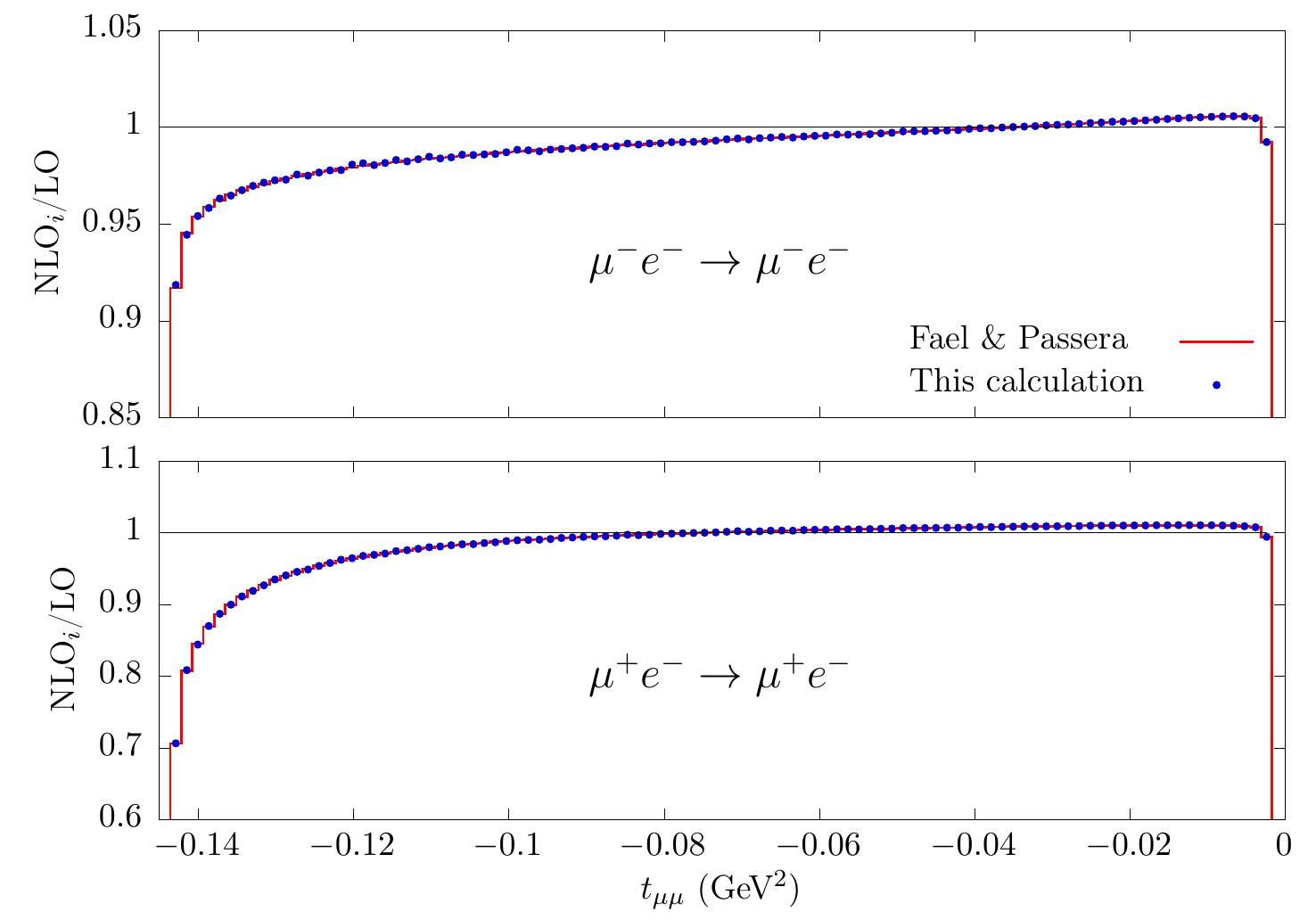}
\end{center}
\caption{\label{Fig:Fig2} The same as in Fig.~\ref{Fig:Fig1}, as
a function  of the squared momentum transfer $t_{\mu\mu} = (p_1 - p_3)^2$.}
\end{figure}

Fig.~\ref{Fig:Fig1} shows a perfect
agreement (within the MC integration statistical error) between the results of the two calculations. 
In Fig.~\ref{Fig:Fig2}, the same level of agreement is observed for
the ratio above as a function of the squared momentum
transfer $t_{\mu\mu} = (p_1 - p_3)^2$. In
Figs.~\ref{Fig:Fig1}-\ref{Fig:Fig2} we used Setup 2 with the further
constraint that $t_{\mu\mu}$ lies in the same range as $t_{ee}$.

\subsection{Integrated cross sections}
\label{sec:integrated}
First, we show numerical results for the radiative corrections at the level of integrated cross section, 
for the different experimental conditions defined above. We focus on the contributions due to 
QED corrections and postpone the discussion of the smaller effects due to electroweak contributions
to Section~\ref{sec:differential}.

\begin{table}[t]
\centering
\begin{tabular}{| c | c | c | c| c|}
\hline
\multicolumn{5}{| c |}{$\bm \mu^{\bm \pm} \bm e^{\bm -} \bm \to \bm\mu^{\bm \pm} \bm e^{\bm -}$}\\
\hline
\multicolumn{3}{| c |}{Cross section} & {Setup 1 \& Setup 3} & {Setup 2 \& Setup 4}\\
\hline
\multicolumn{3}{| c|}{$\sigma_\text{LO}^\text{QED}$} & 1265.060312(7) &  245.038910(1) \\
\hline
\multicolumn{3}{| c|}{$\sigma_\text{LO}^\text{ VP~lep.+top}$} & 1276.876586(9) & 247.946485(1) \\
\hline
\multicolumn{3}{| c|}{$\sigma_\text{LO}^\text{ VP~lep.+top+had.}$} & 1276.903841(10) &  247.965989(2) \\
\hline
\multicolumn{5}{| c |}{$\bm \mu^{\bm +} \bm e^{\bm -} \bm \to \bm \mu^{\bm +} \bm e^{\bm -}$}\\
\hline
Cross section & Setup 1 & Setup 2 & Setup 3 & Setup 4 \\
\hline
$\sigma_\text{NLO}^\text{QED}$ & 1325.217(3) & 255.8437(5) & 1162.447(2) & 222.7714(3) \\
\hline
\multicolumn{5}{| c |}{$\bm \mu^{\bm -} \bm e^{\bm -} \bm \to \bm \mu^{\bm -} \bm e^{\bm -}$}\\
\hline
$\sigma_\text{NLO}^\text{QED}$ & 1323.478(3) & 255.1176(5) & 1161.888(1) & 222.8545(3) \\
\hline
\multicolumn{5}{| c |}{}\\
\hline
\multicolumn{5}{| c |}{$\bm \mu^{\bm \pm} \bm e^{\bm -} \bm \to \bm\mu^{\bm \pm} \bm e^{\bm -}$}\\
\hline
\multicolumn{3}{| c |}{Relative correction} & {Setup 1 \& Setup 3} & {Setup 2 \& Setup 4}\\
\hline
\multicolumn{3}{| c|} {$\delta^\text{ VP}_\text{lep.+top}$} & 0.009340014(8) & 0.011865215(7) \\
\hline
\multicolumn{3}{| c|} {$\delta^\text{VP}_\text{had.}$} & 0.000021345(10) & 0.000078665(9) \\
\hline
\multicolumn{5}{| c |}{$\bm \mu^{\bm +} \bm e^{\bm -} \bm \to \bm \mu^{\bm +} \bm e^{\bm -}$}\\
\hline
Relative correction & Setup 1 & Setup 2 & Setup 3 & Setup 4 \\
\hline
$\delta_\text{NLO}^\text{QED}$ & 0.047552(2)  & 0.044094(2) & -0.081114(1)  &  -0.090874(1) \\
\hline
\multicolumn{5}{| c |}{$\bm \mu^{\bm -} \bm e^{\bm -} \bm \to \bm \mu^{\bm -} \bm e^{\bm -}$}\\
\hline
$\delta_\text{NLO}^\text{QED}$ & 0.046177(2) & 0.041130(2) & -0.081556(1) & -0.090535(1)\\
\hline
\end{tabular}
\caption{\label{tab:tab1}  Cross sections (in $\mu$b) and relative corrections for the 
processes $\mu^+ e^- \to \mu^+ e^-$ and $\mu^- e^- \to \mu^- e^-$, in the four different setup described in the text. 
The digits in parenthesis correspond to the 1$\sigma$ MC error estimate.}
\end{table}

The cross sections quoted in Tab.~\ref{tab:tab1} are defined as follows:
\begin{itemize}

\item $\sigma_\text{LO}^\text{QED}$ and $\sigma_\text{NLO}^\text{QED}$ are the QED cross sections 
at LO and NLO accuracy\footnote{The cross section $\sigma_\text{NLO}^\text{QED}$ includes the contribution 
of purely photonic corrections only, as the gauge-invariant effects of the photon self-energy are evaluated 
separately in $\sigma_\text{LO}^\text{VP~lep.+top}$ and $\sigma_\text{LO}^\text{VP~lep.+top+had.}$. }, respectively;

\item $\sigma_\text{LO}^\text{VP~lep.+top}$ is the LO QED 
cross section including the contribution of the leptonic (+ top-quark) correction to the vacuum polarisation, 
whereas $\sigma_\text{LO}^\text{VP~lep.+top+had.}$ is the LO QED cross section including the 
full correction to the running of $\alpha$.

\end{itemize}

From the above cross section values, we derive the relative corrections shown in  Tab.~\ref{tab:tab1} according to the
following definitions:
\begin{eqnarray}
\delta^\text{ VP}_\text{lep.+top} \, &=& \, \frac{\sigma_\text{LO}^\text{ VP~lep.+top} - \sigma_\text{LO}^\text{QED}}{\sigma_\text{LO}^\text{QED}}
\qquad \qquad 
\delta^\text{ VP}_\text{had.} \, = \, \frac{\sigma_\text{LO}^\text{ VP~ lep.+top+had.} - \sigma_\text{LO}^\text{VP~lep.+top}}{\sigma_\text{LO}^\text{VP~lep.+top}} \nonumber\\
\delta_\text{NLO}^\text{QED} \, &=& \, \frac{\sigma_\text{NLO}^\text{QED} - \sigma_\text{LO}^\text{QED}}{\sigma_\text{LO}^\text{QED}} 
\qquad
\end{eqnarray}

Generally speaking, the LO cross sections in Setup 1 and Setup 3 differ from those of Setup 2 and Setup 4
because of the different electron energy threshold, which implies a different minimum value 
of the squared momentum transfer $t$ allowed by the elastic
kinematics, as remarked above. 
Going from $E_e^\text{min} = 0.2$~GeV to $E_e^\text{min} = 1$~GeV, the elastic cross section is reduced 
by about a factor of five.

More in detail, it can be seen from Tab.~\ref{tab:tab1} that the leptonic (+ top) correction to the
vacuum polarisation, which is dominated by the electron loop contribution, is of the order of
1\% in all the setup. The hadronic correction to the running of the QED coupling 
amounts to $2 \times 10^{-5}$  in Setup 1 and Setup 3 and
$8 \times 10^{-5}$  in Setup 2 and Setup 4. The impact of
$\Delta\alpha_\text{had}(t)$ on the differential cross sections, which
are the relevant observables for the determination of
$a_\mu^{\text{HLO}}$ in the MUonE experiment, is shown in Section~\ref{sec:differential}.

The photonic corrections are positive and rather stable, at the level of 4-5\%, when no acoplanarity cuts are applied. 
They change only by a few per mille when varying the electron energy threshold and are
therefore mildly dependent on the applied electron energy cut. On the other hand, 
they change sign and are enhanced to 8-9\% in the presence of an acoplanarity cut, as a 
consequence of the increasing importance of soft photon emission. Note also that the 
photonic corrections differ by a few per mille (Setup 1 and Setup 2) and at the $10^{-4}$ level 
(Setup 3 and Setup 4) for the processes involving opposite charge muons.
This aspect is discussed in more detailed in the following.

\begin{table}[t]
\centering
\begin{tabular}{| c | c | c | c| c|}
\hline
Relative correction & Setup 1 & Setup 2 & Setup 3 & Setup 4 \\
\hline
\multicolumn{5}{| c |}{$\bm \mu^{\bm +} \bm e^{\bm -} \bm \to \bm \mu^{\bm +}  \bm e^{\bm -}$}\\
\hline
$\delta_\text{NLO}^\text{QED}$ & 0.04755(1) & 0.04409(1) & -0.08111(4) & -0.09087(1) \\
\hline
$\delta_\text{NLO}^\text{electron}$ & 0.04694(2) & 0.04289(1) & -0.08058(3) & -0.08816(1) \\
\hline
$\delta_\text{NLO}^\text{muon}$  & -0.00008(1) & -0.00028(1)  & -0.00078(1) & -0.00256(1)  \\
\hline
$\delta_\text{NLO}^\text{up-down int.}$  & 0.00069(2) & 0.00149(2)  & 0.00025(5)  & -0.00015(2) \\
\hline
\multicolumn{5}{| c |}{$\bm \mu^{\bm -} \bm e^{\bm -} \bm \to \bm \mu^{\bm -}  \bm e^{\bm -}$}\\
\hline
$\delta_\text{NLO}^\text{QED}$ & 0.04618(1) & 0.04114(1) & -0.08154(3)  &  -0.09055(1)\\
\hline
$\delta_\text{NLO}^\text{electron}$ & 0.04694(2)   &   0.04289(1)   & -0.08056(2)  & -0.08817(1) \\
\hline
$\delta_\text{NLO}^\text{muon}$  & -0.00008(1)   & -0.00028(1)  & -0.00078(1) & -0.00256(1)\\
\hline
$\delta_\text{NLO}^\text{up-down int.}$  & -0.00068(2)  & -0.00147(2) & -0.00020(3)  & 0.00017(2)  \\
\hline
\end{tabular}
\caption{\label{tab:tab2}  Relative corrections for 
the gauge-invariant subsets of QED corrections to the 
processes $\mu^+ e^- \to \mu^+ e^-$ and $\mu^- e^- \to \mu^- e^-$, in the four different setup described in the text. 
The digits in parenthesis correspond to the 1$\sigma$ MC error estimate.}
\end{table}

Within the full set of NLO QED corrections, there are three gauge-invariant subsets of 
photonic contributions corresponding to:
\begin{itemize}

\item virtual and real corrections along the electron line;

\item virtual and real corrections along the muon line;

\item up-down interference connecting the electron and muon line, due
  to box contributions and up-down interference of real photon radiation.

\end{itemize}

The relative contributions associated to the above classes of
correction are shown in Tab.~\ref{tab:tab2},
where it can be noticed that the corrections due
to electron and muon radiation 
do not change when considering the processes initiated by opposite charge muons. 
On the other hand, the up-down interference corrections have the same size but opposite sign, as 
can be expected from the different charge pattern occurring 
in the interference/box contribution to the two processes. This explains the 
difference between the NLO QED corrections to the processes with positive and negative muons.

From Tab.~\ref{tab:tab2}, one can also see that for  both the $\mu^+ e^- \to \mu^+ e^-$ and $\mu^- e^- \to \mu^- e^-$ process
the overall (some per cent) corrections are largely dominated by the contribution due to electron radiation. 
In all the setup, the latter provides more than 95\% of the whole correction. The contributions 
due to muon radiation and electron-muon interference share the remaining part of the
full correction and lie in the range between $10^{-4}$
and $10^{-3}$. In particular, for Setup 1 and Setup 2, the next-to-dominant contribution 
is given by the up-down interference corrections, whereas in Setup 3 and Setup 4 muon radiation 
exceeds the effect of the NLO up-down interference. A more detailed analysis of the
contributions due to the different classes of corrections for the
differential distributions is given in the next Section.

\subsection{Differential cross sections}
\label{sec:differential}

More than the integrated cross sections, various distributions are relevant for 
the newly proposed MUonE experiment. We mainly focus on the $\mu^+ e^- \to \mu^+ e^-$ process 
because of the larger fraction of positive muons provided by the CERN M2 beam. 
Particular attention is paid to the differential cross sections as functions of the angular variables,
as the proposed experimental arrangement of the MUonE
experiment is primarily designed to measure the angles precisely~\cite{Abbiendi:2016xup}. 

\begin{figure}[hbtp]
\begin{center}
\includegraphics[width=0.7\textwidth]{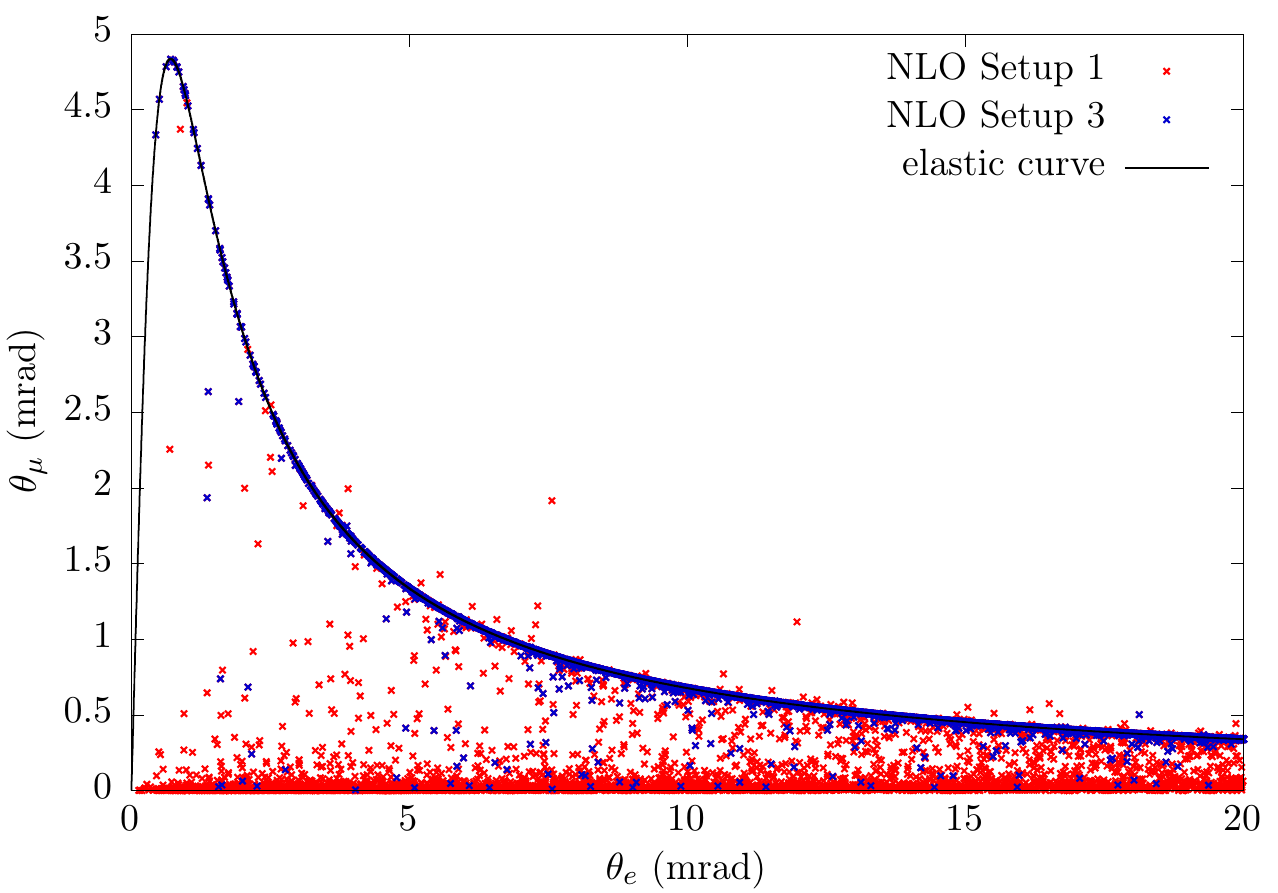}
\end{center}
\caption{\label{Fig:Fig3} The correlation between the electron scattering angle $\theta_e$ and muon 
scattering angle $\theta_\mu$ for the $\mu^+ e^- \to \mu^+ e^-$ process at LO (elastic curve) and 
NLO QED, for the selection criteria 1 and 3 defined in the text.}
\end{figure}

\begin{figure}[h]
\begin{center}
\includegraphics[width=0.5\textwidth]{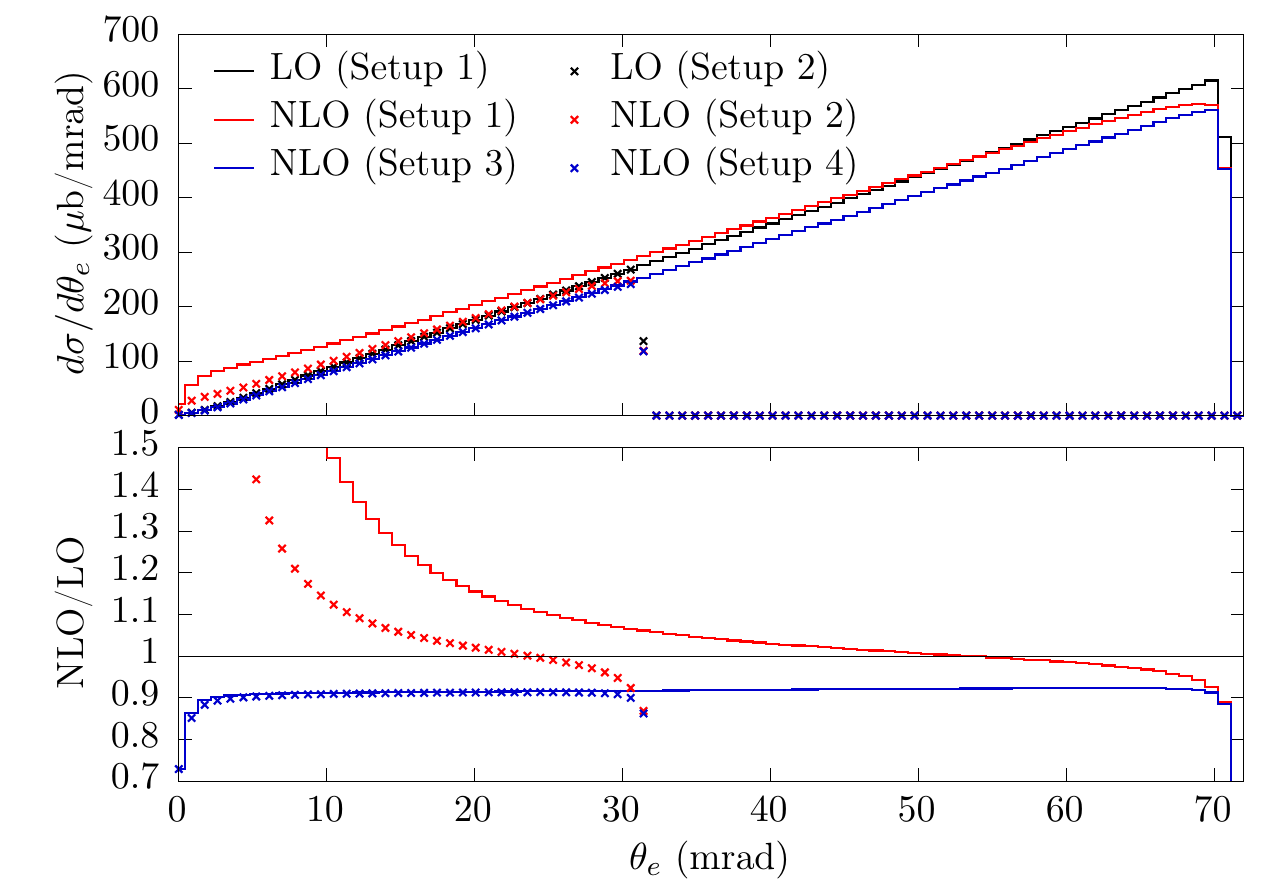}~~\includegraphics[width=0.5\textwidth]{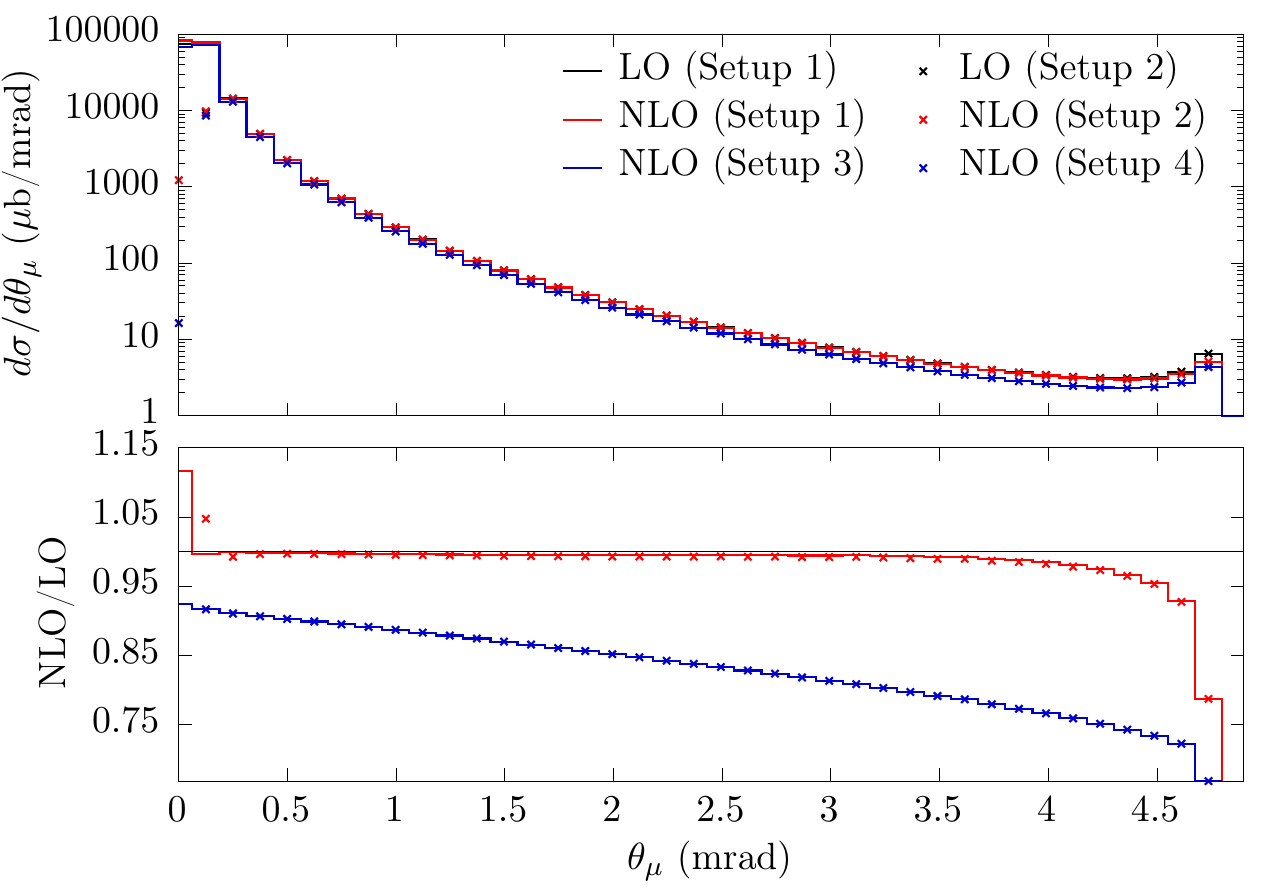}
\end{center}
\caption{\label{Fig:Fig4} Left plot: the LO and NLO QED cross sections of the 
$\mu^+ e^- \to \mu^+ e^- $ process, as a function of the electron scattering angle, for the 
different setup described in the text (upper panel); 
relative NLO QED corrections (lower panel). Right plot: the same as in the left plot, as a function of the 
muon scattering angle.}
\end{figure}

In Fig.~\ref{Fig:Fig3} we compare the correlation, {obtained with a
sample of MC events}, between the scattering angles of the outgoing electron and muon
at LO and NLO, for the Setup 1 and Setup 3 defined above. It can be noticed that, in the 
absence of an acoplanarity cut (Setup 1), the correlation present at LO 
(elastic curve) is largely modified by the presence of events at relatively small muon angles, which 
originate from the bremsstrahlung process $\mu^+ e^- \to \mu^+ e^- + \gamma$.  However, the tight 
acoplanarity cut (Setup 3) turns out to be effective in getting rid of
most of these radiative events, thus isolating the elastic
{correlation curve}.
Other and more sophisticated elasticity conditions~\cite{Amendolia:1986wj} could be applied on the experimental 
side and easily taken into account at the simulation level.

\begin{figure}[h]
\begin{center}
\includegraphics[width=0.5\textwidth]{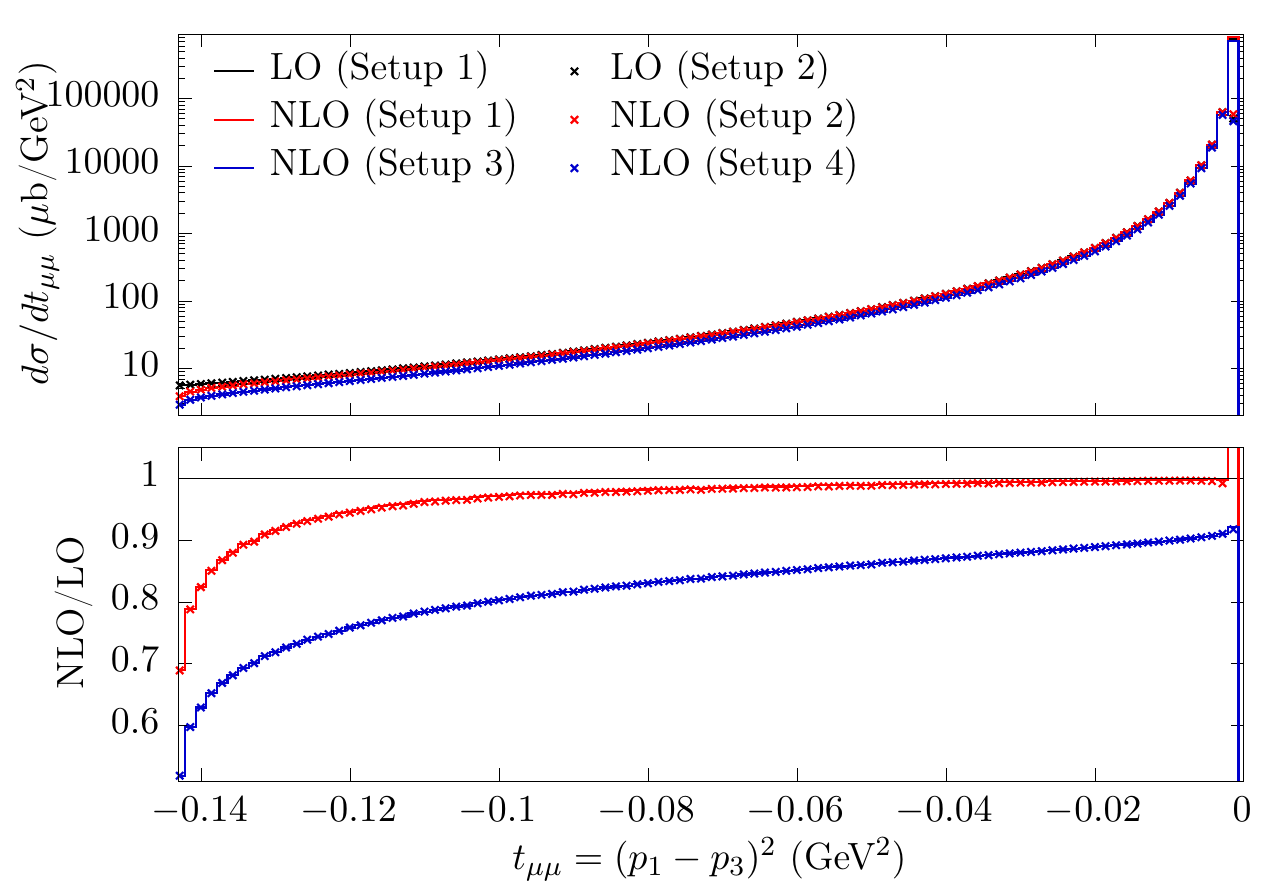}~~\includegraphics[width=0.5\textwidth]{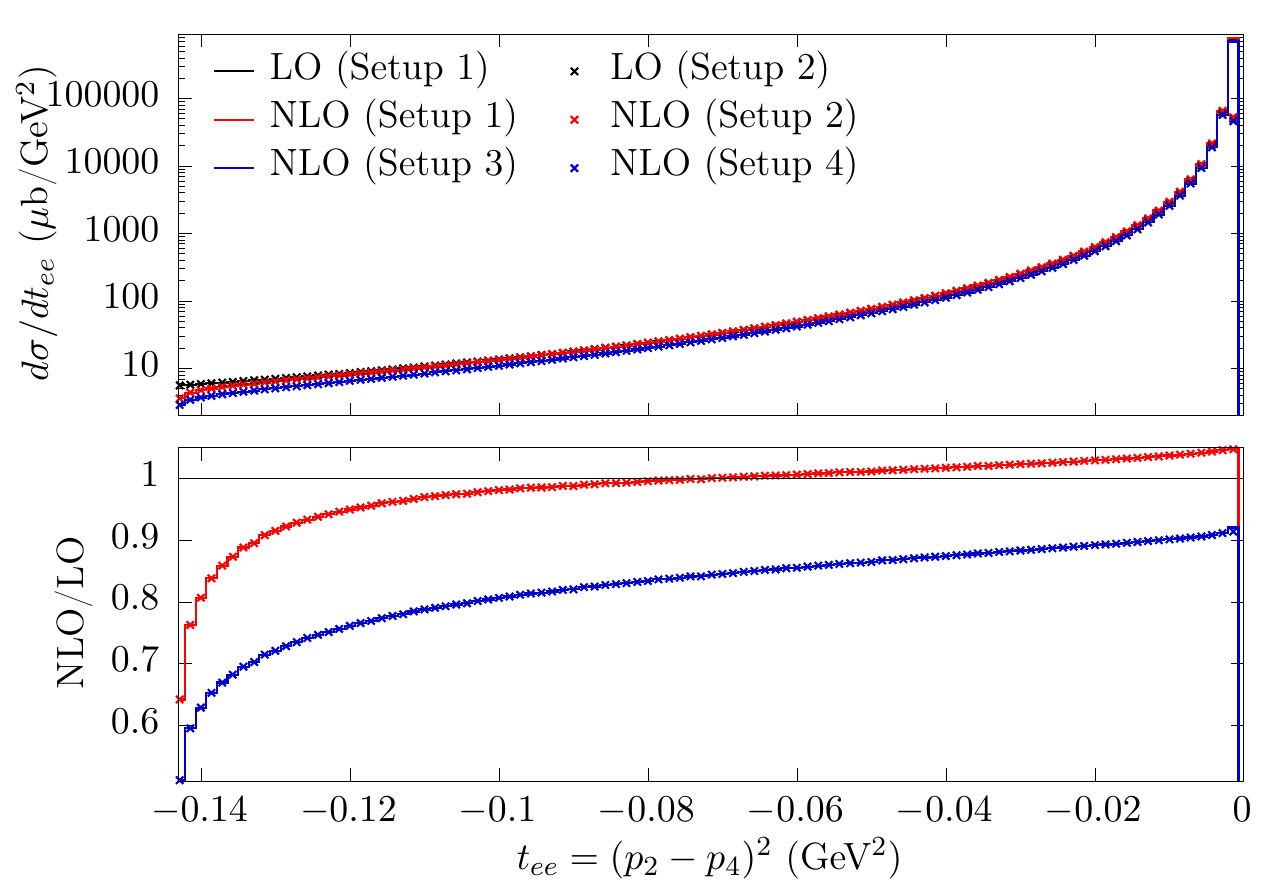}
\end{center}
\caption{\label{Fig:Fig5} The same as Fig.~\ref{Fig:Fig4} for the cross sections as a function of the squared momentum transfer $t_{\mu\mu}$ (left plot) and $t_{ee}$ 
(right plot).}
\end{figure}

\begin{figure}[h]
\begin{center}
\includegraphics[width=0.5\textwidth]{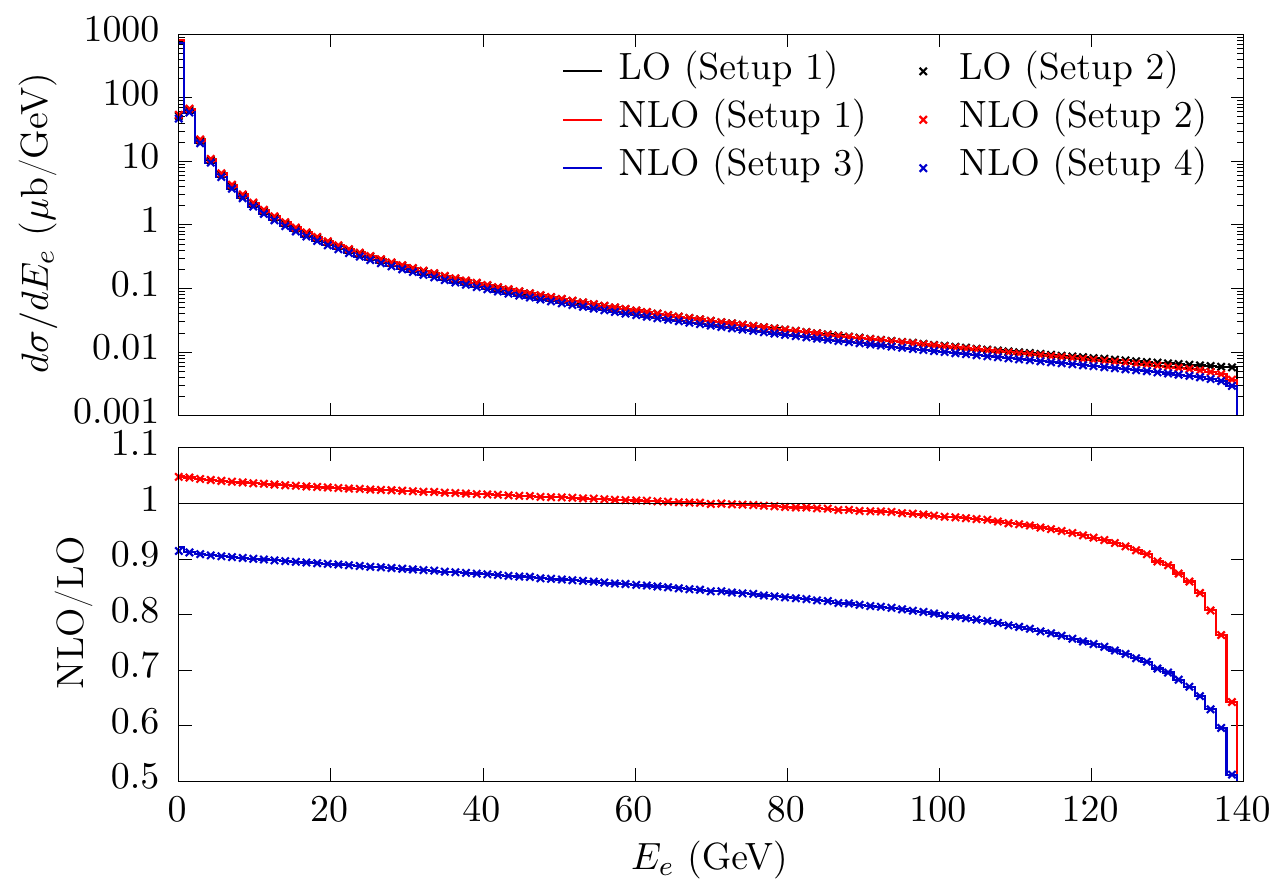}~~\includegraphics[width=0.5\textwidth]{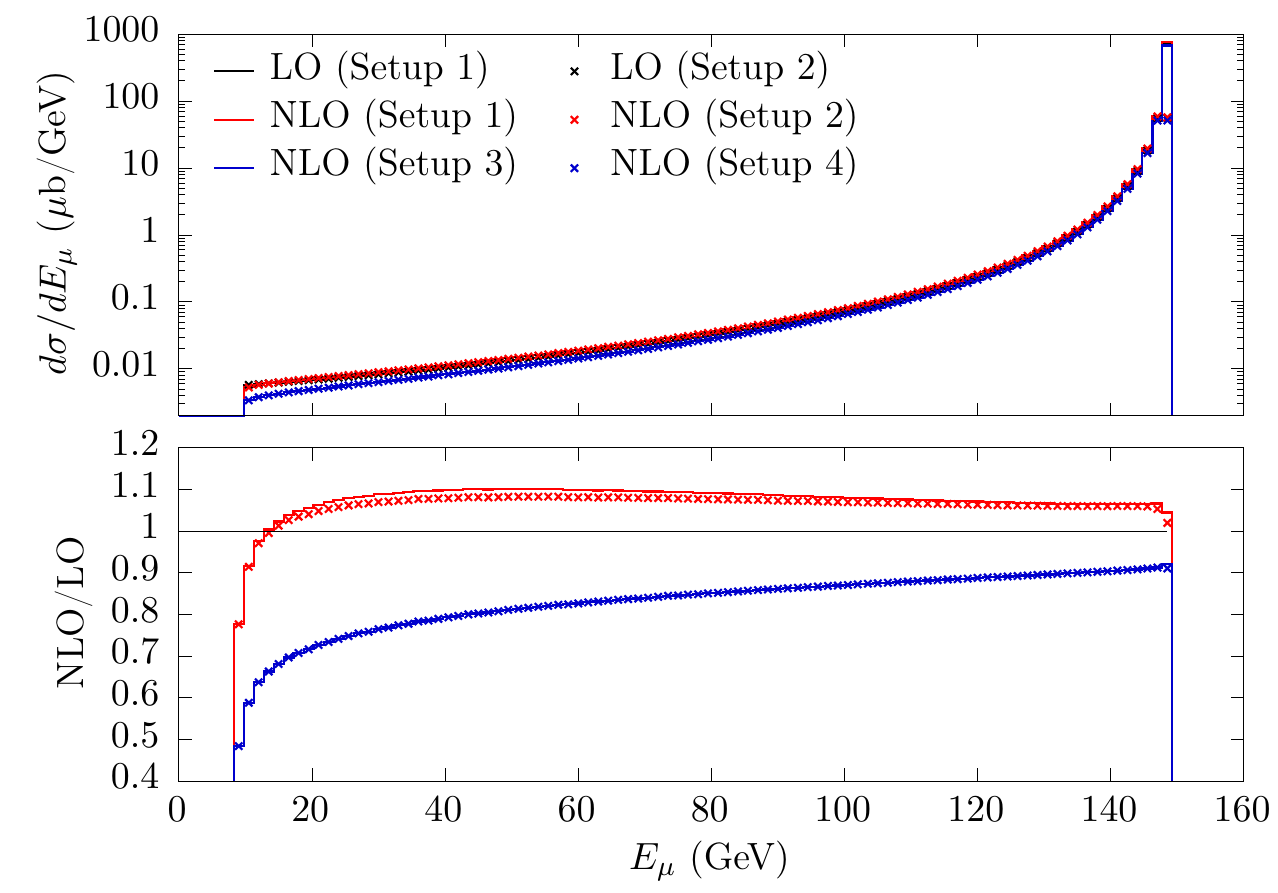}
\end{center}
\caption{\label{Fig:Fig6} The same as Fig.~\ref{Fig:Fig4} for the cross sections as a function of the electron (left plot) and muon (right plot) energy.}
\end{figure}

In Figs.~\ref{Fig:Fig4}-\ref{Fig:Fig6} we show the impact of the NLO QED (purely photonic) corrections on a 
number of differential cross sections of the process $\mu^+ e^- \to \mu^+ e^-$ process, in order to study
how the different observables are affected by the radiative corrections and applied cuts.

In Fig.~\ref{Fig:Fig4} the effect of the QED corrections under the different event selection criteria is shown 
for the distribution of the electron (left plot) and the muon scattering angles (right plot). 
It can be clearly seen that, when no elasticity cuts are applied, the electron angle distribution is significantly 
affected by the QED contributions 
in the region of small scattering angles, which is the part of the cross section of main interest for the
extraction of $\Delta\alpha_\text{had}(t)$.
This strongly varying distortion, which
corresponds to an enhancement of events at relatively small
muon angles, can be understood as an effect
induced on the electron by the radiative processes $\mu^+ e^- \to \mu^+ e^- + \gamma$, as
already remarked for Fig.~\ref{Fig:Fig3}.
However, this effect largely disappears if an acoplanarity cut
is imposed, thus giving rise to a flat and less significant correction of the order of 10\%. 
It is worth noting that the same does not occur for the $\theta_\mu$ distribution. In the absence of an acoplanarity cut,
the latter distribution receives a flat and moderate contribution from the radiative corrections 
(independently of the electron energy threshold), with the 
exception of the upper hedge where the QED contributions can reach the 20\% level. In the presence of 
an elasticity condition, the corrections to the muon scattering angle distribution become enhanced and more 
varying, as a consequence of the more pronounced importance of soft photon radiation.

In Fig.~\ref{Fig:Fig5} we show the contribution of the photonic
corrections to the differential cross sections as functions
of the squared momentum transfer $t_{\mu\mu}$ (left plot) and $t_{ee}$ (right plot). We can notice that, for the 
situation of acceptance cuts only (Setup 1 and Setup 2), there are small and constant corrections for
values $|t_{\mu\mu,ee}| \lesssim 0.1~\text{GeV}^2$.
Again, in the presence of an acoplanarity cut, the corrections become larger because of the growing 
importance of soft photon emission and vary in the range between 10\% and 40\%.

Concerning the energy of the outgoing leptons, we show in Fig.~\ref{Fig:Fig6} the energy spectrum 
of the electron (left plot) and of the muon (right plot),
respectively. Generally speaking, it can be noticed 
that the corrections are to a large extent independent of the applied electron energy cut, as they are
practically the same for Setup 1 and Setup 2. For these selection criteria, the corrections are slowly 
varying in almost all the range but they grow up to some tens of per cent in the limit of high electron 
energy and small muon energy. This behaviour can be ascribed to the dominant r\^ole played 
by the radiation emitted by the electron leg, which turns out to be emphasised in the above
regions by the contribution of soft photons. However, it is worth noting that the cross sections
corresponding to the above kinematical limits are pretty small. When an elasticity condition is 
imposed, the QED corrections to the lepton energies lie in the range between 10\% and 50\%.

\begin{figure}[t]
\begin{center}
\includegraphics[width=0.5\textwidth]{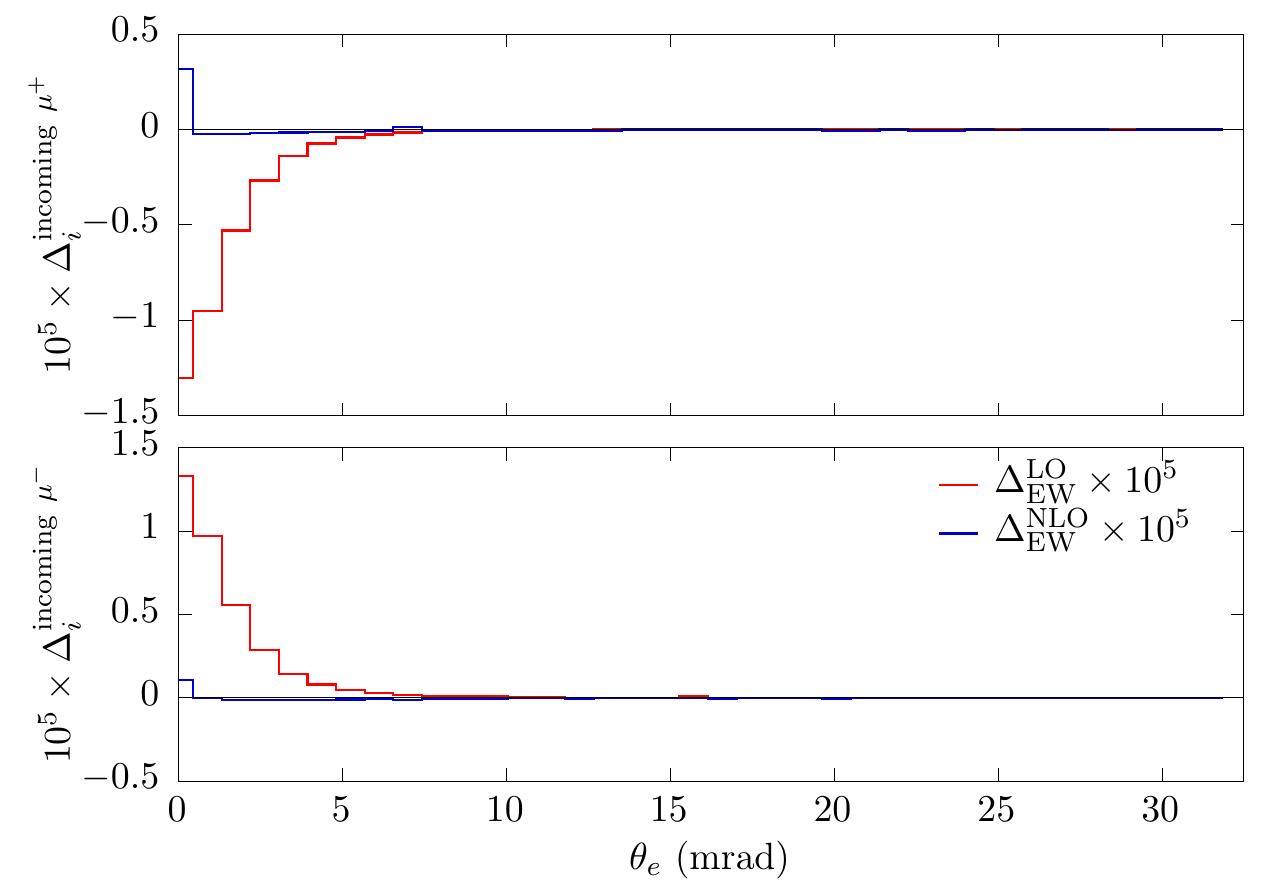}~~\includegraphics[width=0.5\textwidth]{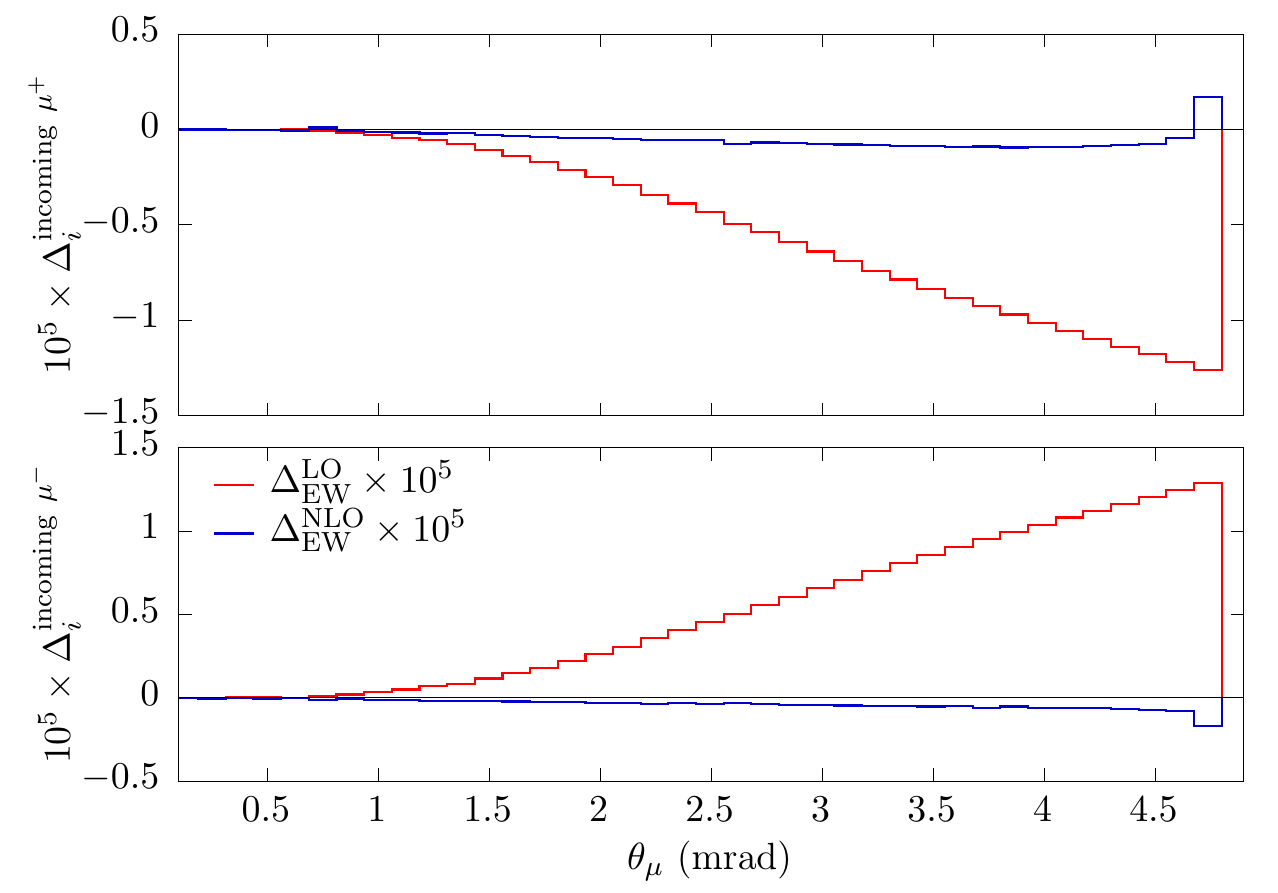}
\end{center}
\caption{\label{Fig:Fig7}  Left plot: the LO and NLO electroweak contributions to the 
cross section of the process $\mu^+ e^- \to \mu^+ e^-$ (upper panel) and of the 
process $\mu^- e^- \to \mu^- e^-$ (lower panel), as a function of the electron scattering angle. Right plot: 
the same as in the left plot as a function of the muon scattering angle. 
The results refer to Setup 2 described in the text.}
\end{figure}

\begin{figure}[t]
\begin{center}
\includegraphics[width=0.5\textwidth]{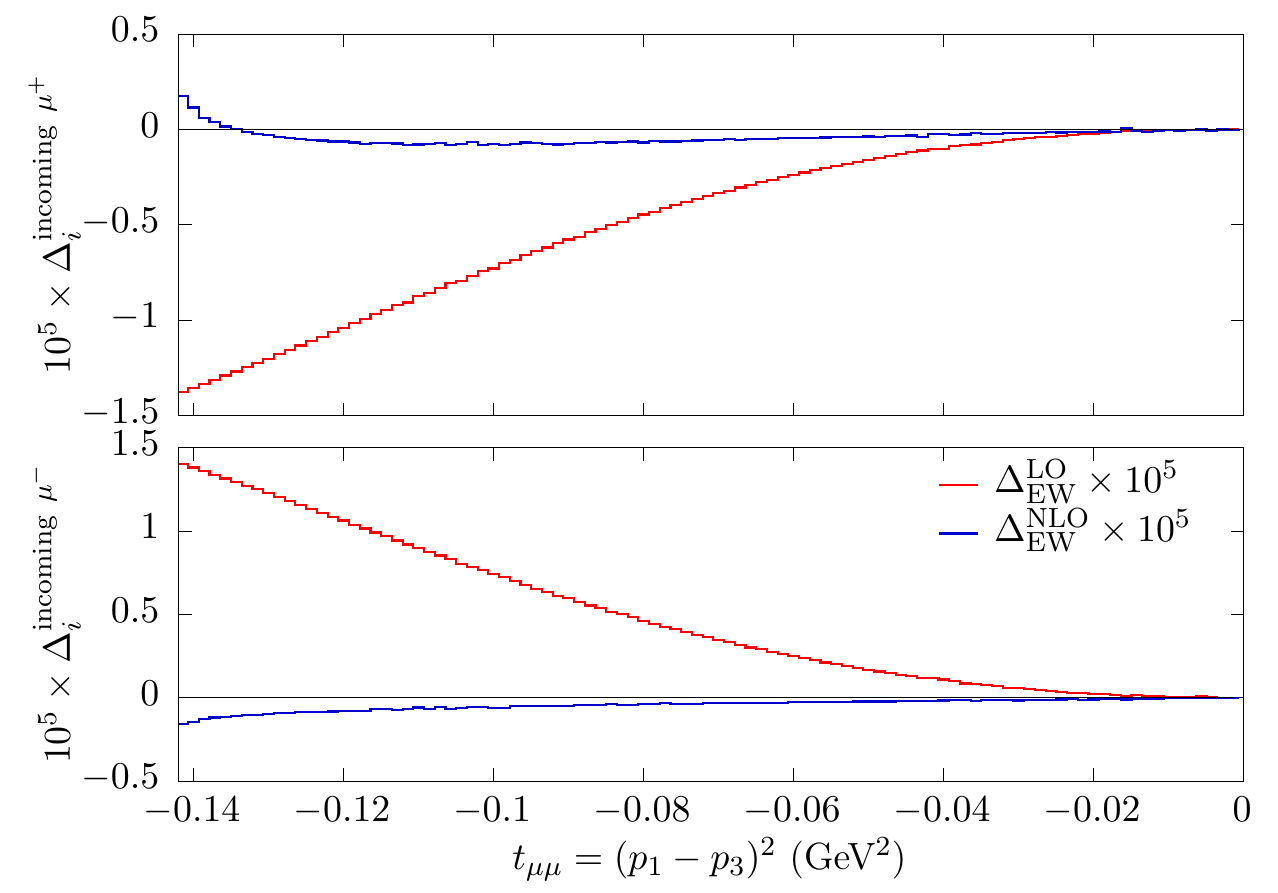}~~\includegraphics[width=0.5\textwidth]{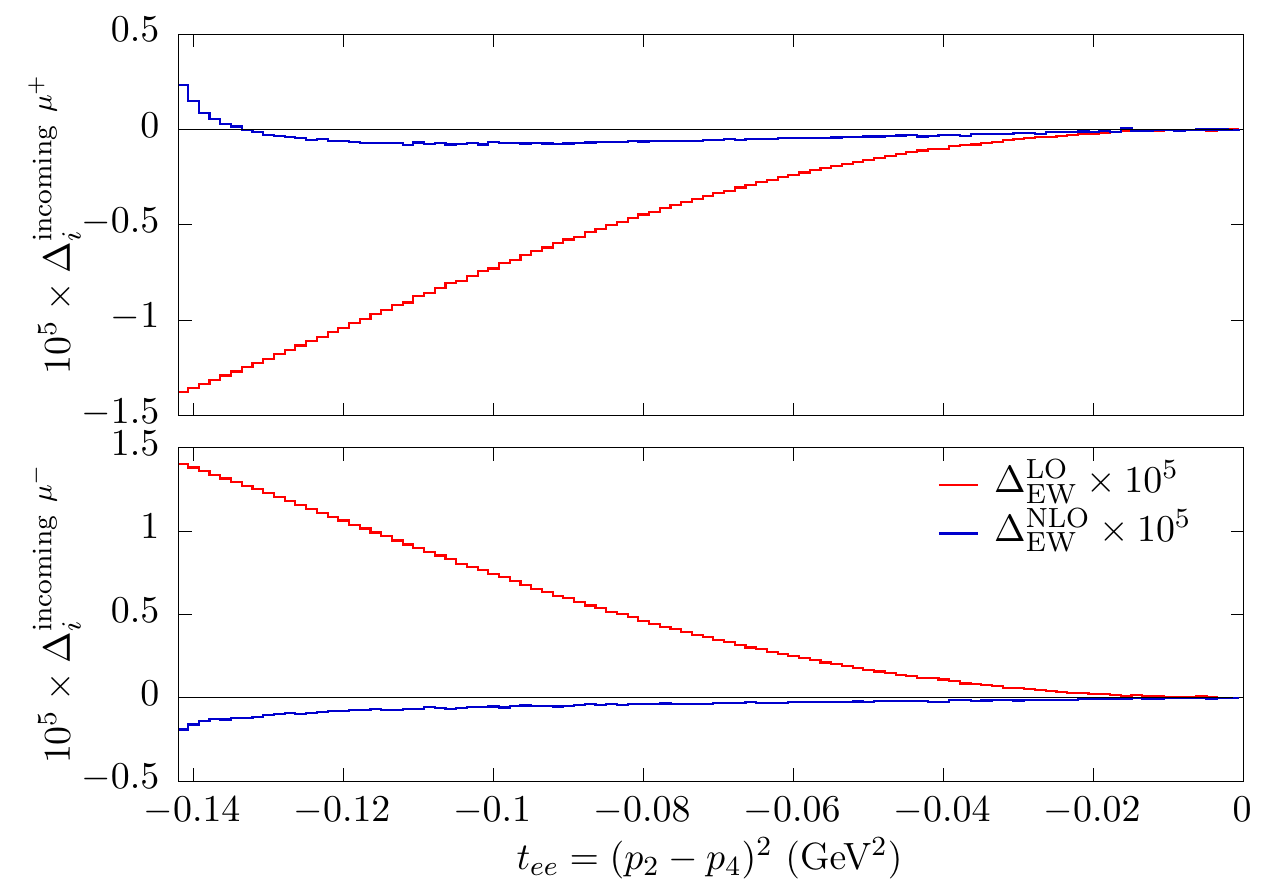}
\end{center}
\caption{\label{Fig:Fig8} The same as Fig.~\ref{Fig:Fig7} for the cross sections 
of the processes $\mu^\pm e^- \to \mu^\pm e^-$, as functions of the squared momentum transfer $t_{\mu\mu}$ (left plot) and $t_{ee}$ 
(right plot).}
\end{figure}

\begin{figure}[ht]
\begin{center}
\includegraphics[width=0.5\textwidth]{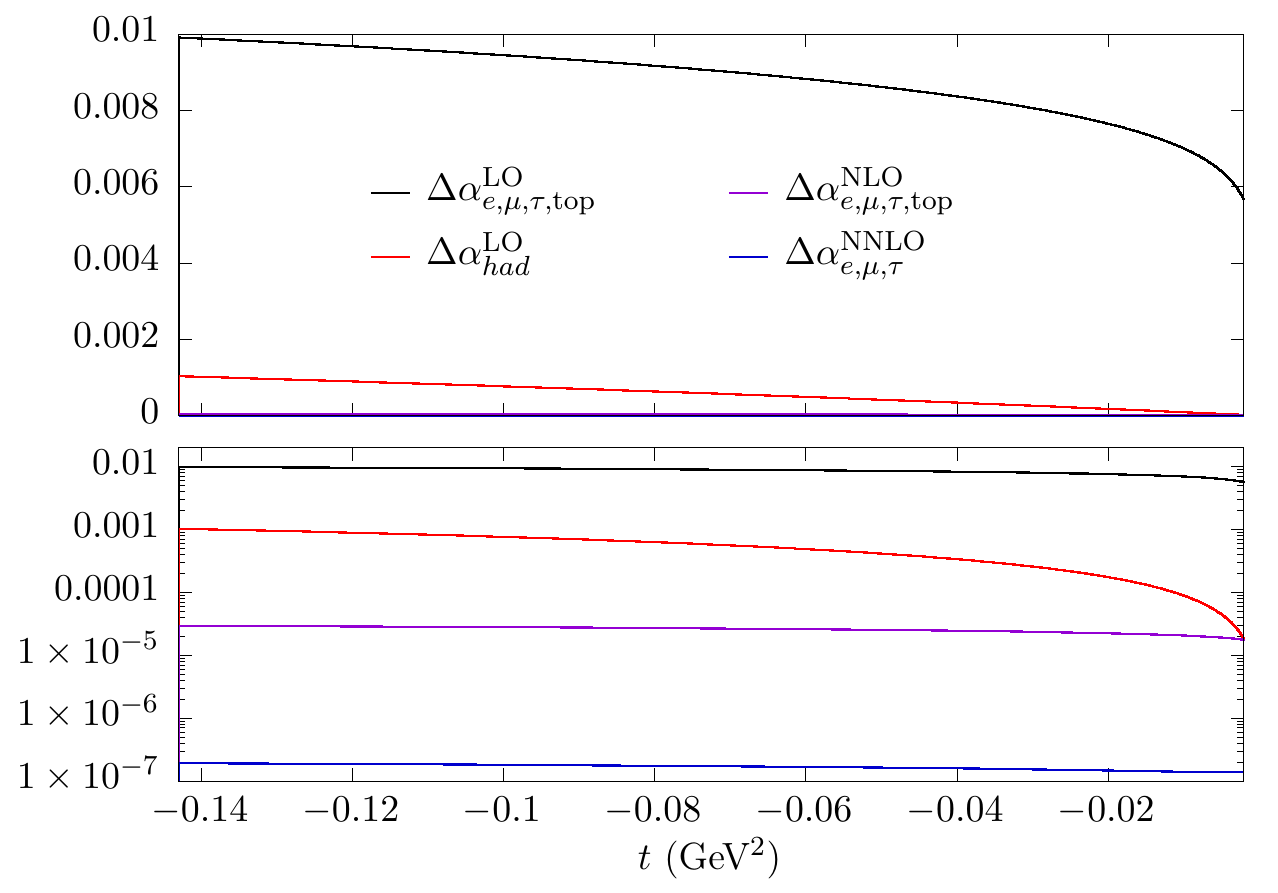}~~\includegraphics[width=0.5\textwidth]{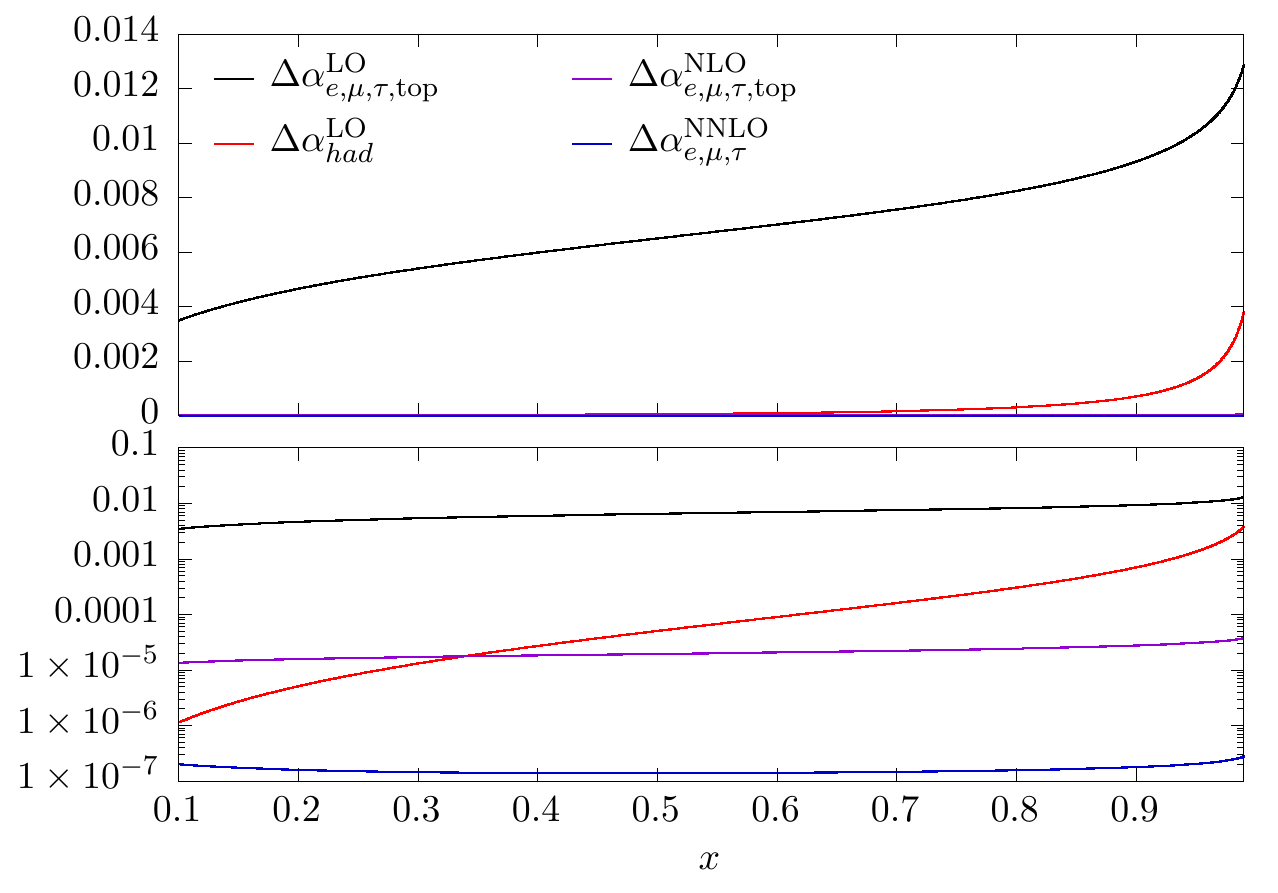}
\end{center}
\caption{\label{Fig:Fig9} Left plot: the leptonic (+ top) correction and the hadronic correction to the running of 
$\alpha_\text{QED}$, as a function of the squared momentum transfer $t$ in linear (upper panel) and log scale (lower panel). 
Right plot: the same as in the left plot as a function of the Feynman variable $x$.}
\end{figure}

\begin{figure}[ht]
\begin{center}
\includegraphics[width=0.5\textwidth]{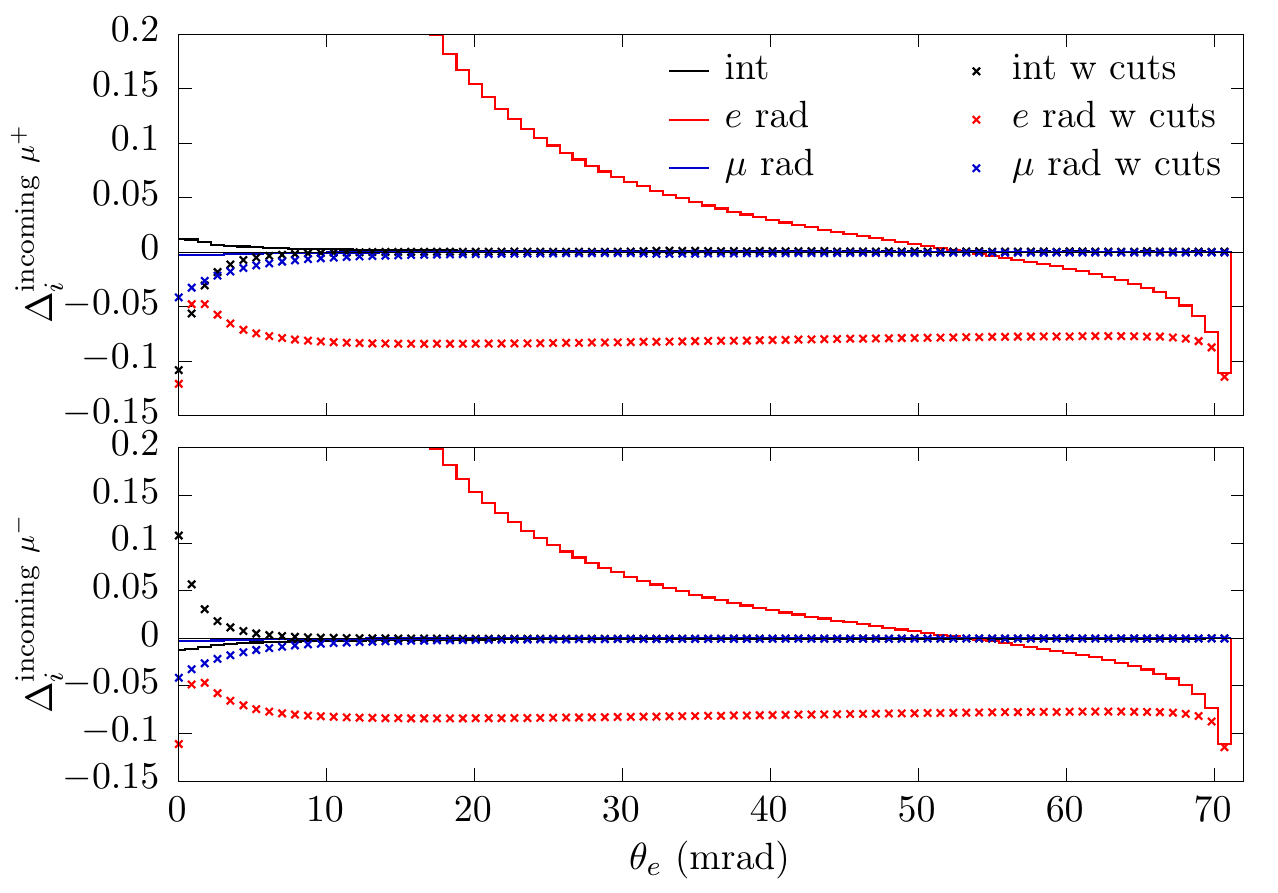}~~\includegraphics[width=0.5\textwidth]{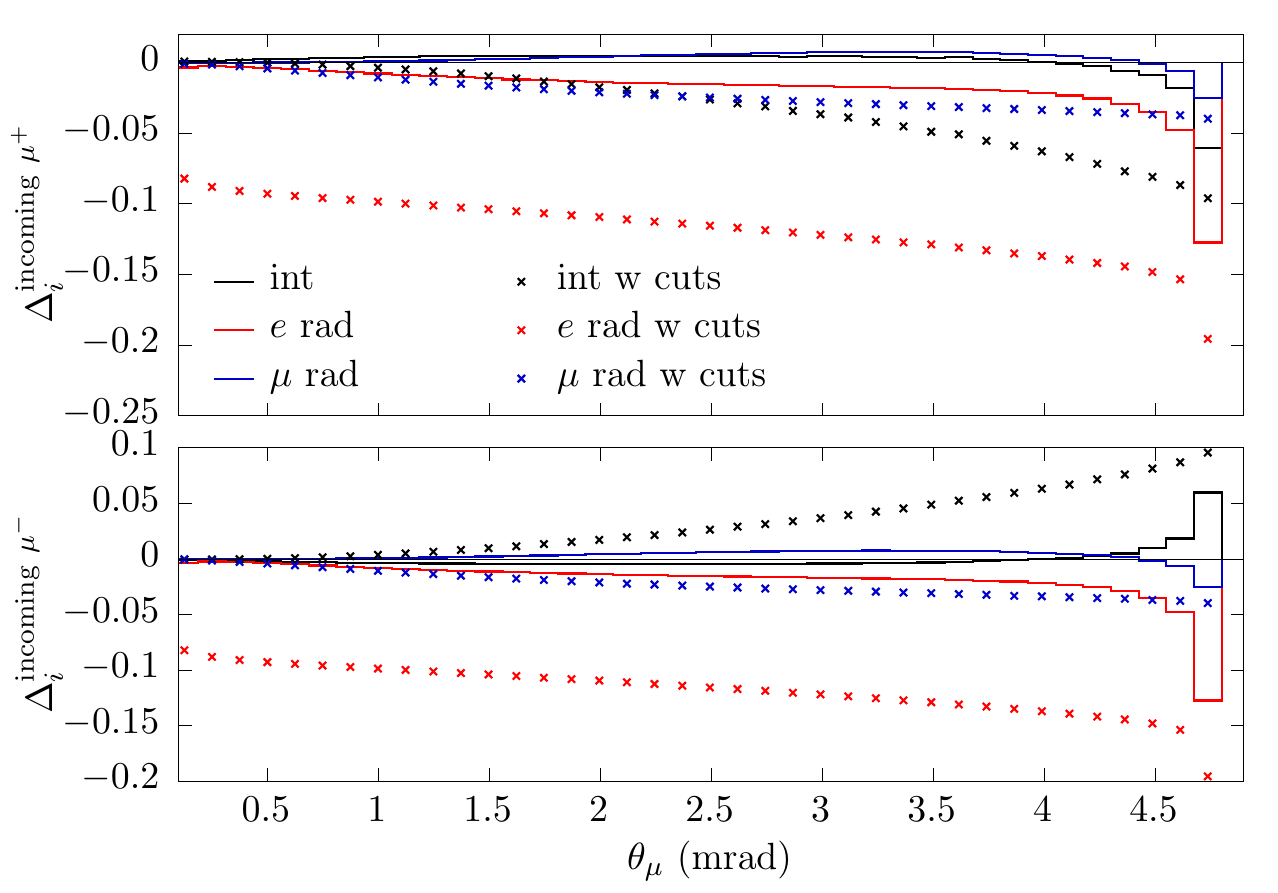}
\end{center}
\caption{\label{Fig:Fig10} Left plot: the contribution of the QED gauge-invariant subsets 
of photonic corrections to the cross section of the process $\mu^+ e^- \to \mu^+ e^-$ (upper panel) and of the 
process $\mu^- e^- \to \mu^- e^-$ (lower panel), as a function of the electron scattering angle. Right plot: 
the same as in the left plot as a function of the muon scattering angle. The results refer to 
Setup 1 and Setup 3 described in the text.}
\end{figure}

\begin{figure}[ht]
\begin{center}
\includegraphics[width=0.5\textwidth]{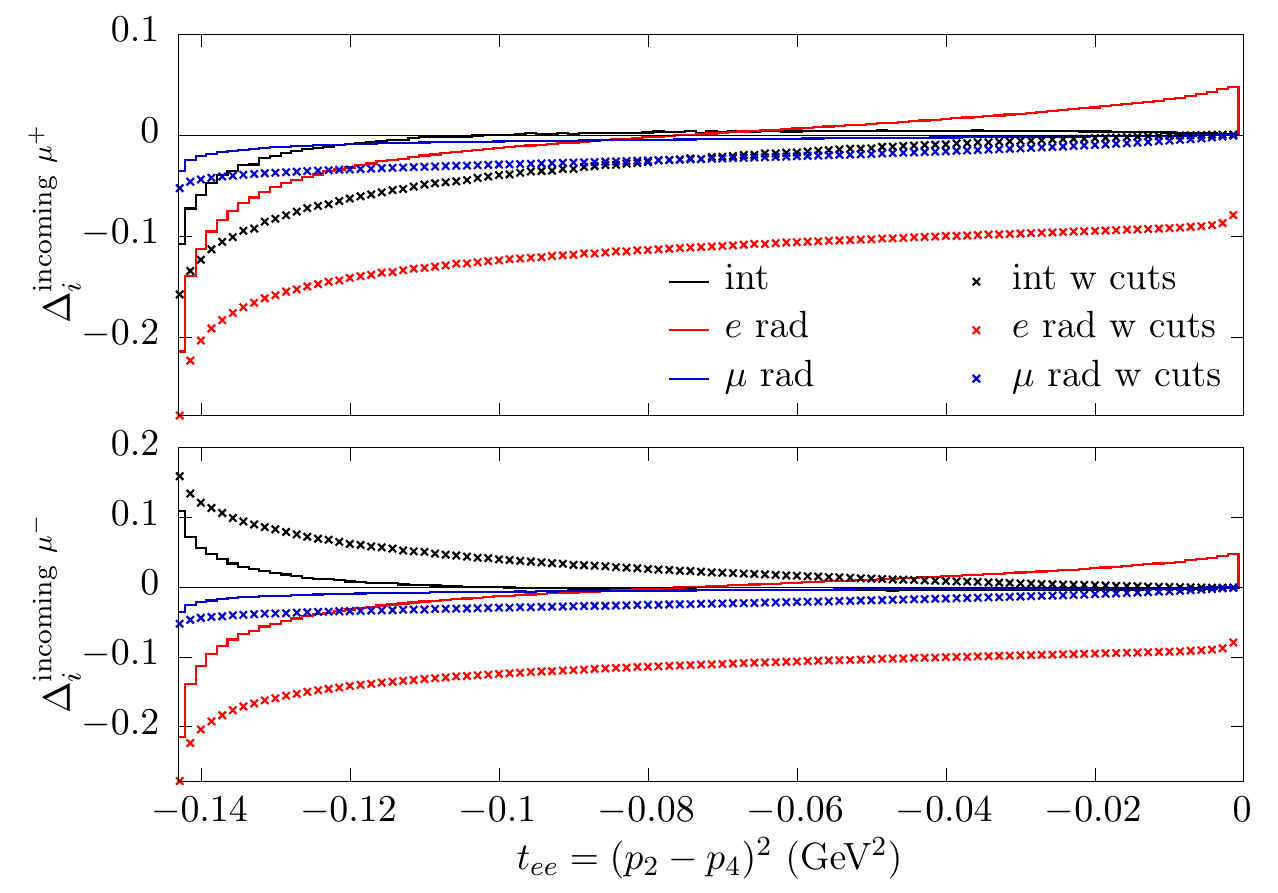}~~\includegraphics[width=0.5\textwidth]{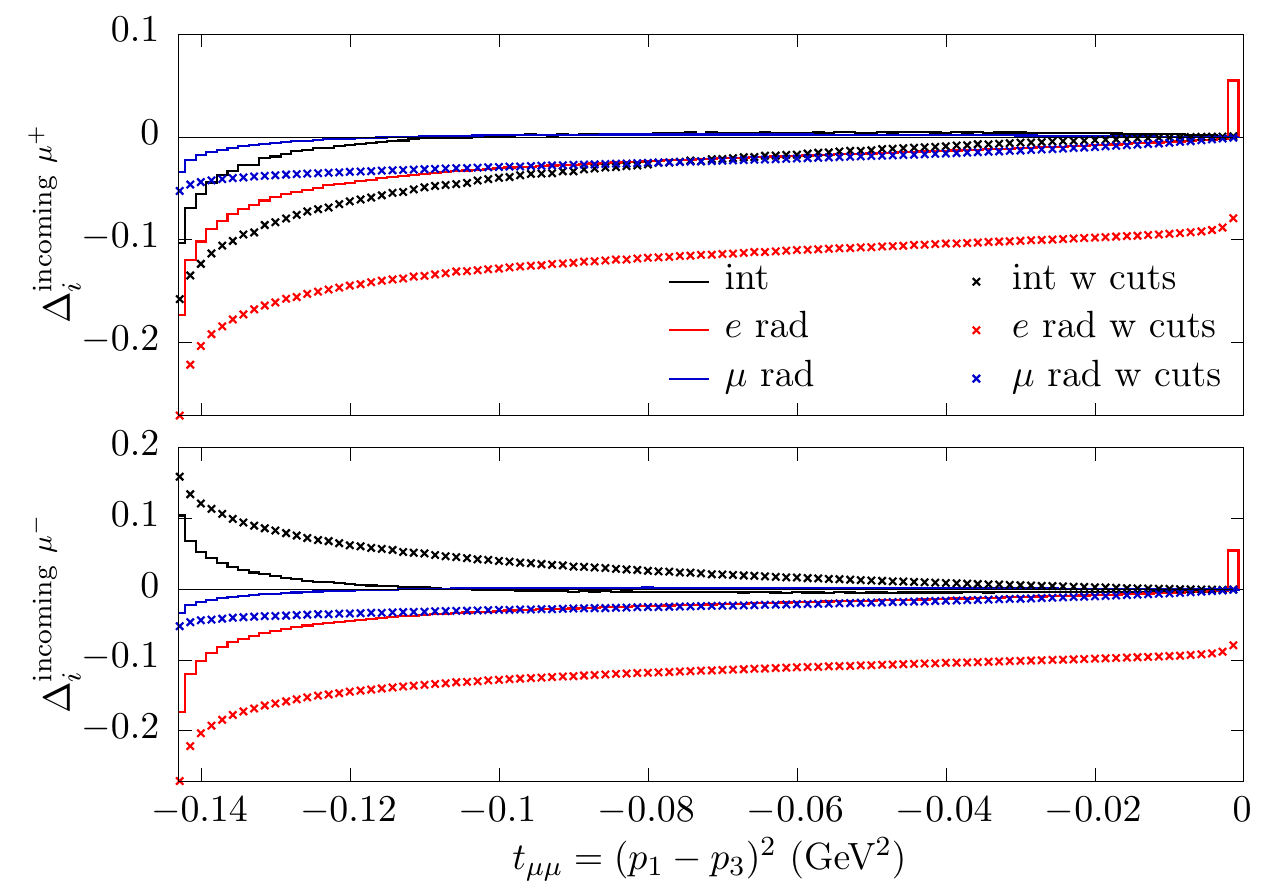}
\end{center}
\caption{\label{Fig:Fig11} The same as Fig.~\ref{Fig:Fig10} for the cross sections 
 for the cross sections 
of the processes $\mu^\pm e^- \to \mu^\pm e^-$, as a function of the squared momentum transfer $t_{\mu\mu}$ (left plot) and $t_{ee}$ 
(right plot).}
\end{figure}

We come now to the discussion of the effects due to the electroweak contributions 
and vacuum polarisation. We also examine in detail the impact of the 
gauge-invariant subsets of photonic corrections at the level of distributions, thus
completing the analysis presented in Section~\ref{sec:integrated} for the
integrated cross sections. We provide results which are valid for the processes 
$\mu^\pm e^- \to \mu^\pm e^-$ and are given in Figs.~\ref{Fig:Fig7}-\ref{Fig:Fig11}.

In Fig.~\ref{Fig:Fig7} the size of the electroweak contributions at LO and NLO accuracy is shown 
for the cross sections $d\sigma / d\theta_e$ (left plot) and $d\sigma / d\theta_\mu$ (right plot) of the 
processes $\mu^\pm e^- \to \mu^\pm e^-$. The same effects are shown in Fig.~\ref{Fig:Fig8}
for the cross sections as a function of the squared momentum transfer $t_{\mu\mu}$ 
and $t_{ee}$. In order to consistently disentangle pure electroweak
effects, we define $\Delta_\text{EW}^\text{LO}$ and $\Delta_\text{EW}^\text{NLO}$ as follows
\begin{eqnarray*}
  \Delta_\text{EW}^\text{LO}=\frac{d\sigma_\text{EW}^\text{LO}-d\sigma_\text{QED}^\text{LO}}{d\sigma_\text{QED}^\text{LO}}
  \qquad\qquad
  \Delta_\text{EW}^\text{NLO}=\frac{\left(d\sigma_\text{EW}^\text{NLO}-
    d\sigma_\text{EW}^\text{LO}\right) - \left(d\sigma_\text{QED}^\text{NLO}-d\sigma_\text{QED}^\text{LO}\right)}{d\sigma_\text{QED}^\text{NLO}}\end{eqnarray*}

From Figs.~\ref{Fig:Fig7}-\ref{Fig:Fig8}, it can be noticed that the LO electroweak contributions to the 
observables of the  $\mu^+ e^- \to \mu^+ e^-$ process (upper panels) have a different shape and
sign with respect to the same observables of the
 $\mu^- e^- \to \mu^- e^-$ process (lower panels). 
This asymmetric behaviour is a 
 parity-violation effect induced by the presence of the $Z$-exchange contribution 
 in the LO diagrams. From  Fig.~\ref{Fig:Fig7}, one can also see that 
 the $\gamma$-$Z$ tree-level contributions are at the $10^{-5}$ level
 for small electron angles and relatively large muon angles. 
 Effects of the same order are observed in Fig.~\ref{Fig:Fig8} for the cross sections as a function of the 
 momentum transfer $t_{\mu\mu}$ and $t_{ee}$ for $|t_{\mu\mu,ee}|$ values larger than about $0.1~\text{GeV}^2$. 
 In view of the target accuracy of the experiment, we can conclude that the LO electroweak 
 contributions must be taken into account in any calculation aiming at
 a 10ppm precision. 
 Concerning the impact of the NLO electroweak corrections, it can be seen that they are 
 almost flat and
 negligible for all the differential cross sections, their contribution being 
 well below $10^{-5}$.
 We checked that also for the electron and muon energy the LO electroweak 
 contributions can reach the 10ppm level, the one-loop electroweak corrections being always
 negligible. 
 
According to the proposed strategy of the MUonE experiment, the leading hadronic contribution to 
the muon $g-2$ can be derived after subtracting the leptonic correction to the running 
of the QED coupling as extracted from the data~\cite{Abbiendi:2016xup}. Therefore, 
it is important to establish the accuracy in the calculation of the leptonic contributions 
to the vacuum polarisation. 
To this end, we show in Fig.~\ref{Fig:Fig9} the behaviour of the leptonic corrections 
to the running of $\alpha$ at the different
perturbative orders~\cite{Steinhauser:1998rq}, 
as a function of the squared momentum transfer $t$ (left plot) and 
of the Feynman-parameter variable $x$ (right plot).
As a reference, also the hadronic contribution is shown, as calculated using the dispersive 
approach available in the {\tt hadr5n16.f} routine~\cite{Jegerlehner:2017zsb}. 
The results as a function of $x$ are shown since the contribution $a_\mu^\text{HLO}$
can be obtained in the space-like approach as a one-dimensional integral over the 
variable $x$ involving the hadronic correction to $\alpha_\text{QED}$ evaluated at
$t (x) = x^2 m_\mu^2 / (x - 1)$~\cite{Calame:2015fva,Abbiendi:2016xup}.
 As can be seen, the leptonic corrections at LO (one-loop approximation) vary from 1\% 
at large $|t|$ and $x$ values to some per mille at small $|t|$ and $x$. 
The NLO (two-loop accuracy) and NNLO (three-loop approximation) leptonic corrections 
are of the order of $10^{-5}$ and $10^{-7}$, respectively. Therefore, the present knowledge
of the above corrections does not provide a limitation to the proposed experimental strategy. 
The hadronic contribution varies from 0.1\% at relatively large $t$
and $x$ values, which define the ``signal region'' of the experiment, to the $10^{-5} - 10^{-6}$ level at small $t$ and $x$, where 
the cross section of the process can be used as normalisation.

An examination of the different sources of photonic corrections is shown for 
the $d\sigma / d\theta_e$ and for $d\sigma / d\theta_\mu$ in Fig.~\ref{Fig:Fig10}
and the $d\sigma / d t_{\mu\mu}$ and $d\sigma / d t_{ee}$ in Fig.~\ref{Fig:Fig11}. 
The upper(lower) panels refer to the $\mu^+ e^- \to \mu^+ e^-$($\mu^- e^- \to \mu^- e^-$) 
process. This analysis is performed in order to understand how the different gauge-invariant 
classes contribute to the overall QED corrections previously discussed. 
The main message that can be drawn by inspection of 
Fig.~\ref{Fig:Fig10} and Fig.~\ref{Fig:Fig11} is that the overall QED correction 
over the full range is, in general, the result 
of a delicate interplay between the various sources of radiation. A further important 
and general remark is that the corrections due to up-down interference are of opposite sign 
for the two processes, which is the origin of the different overall correction for the processes
involving positive and negative muons. Moreover, one can notice that the interference contribution 
plays a significant r\^ole in the region of small electron scattering angles, i.e. for large
$|t_{\mu\mu,ee}|$ values. This behaviour has to be ascribed to the presence
 in the up-down interference
of logarithmic (and squared logarithmic) angular contributions of the type $\ln (u / t)$, 
which become potentially enhanced when either $t$ or $u$ are small.

More in detail, one can see from Fig.~\ref{Fig:Fig10} that, in the presence of acceptance cuts 
only, the correction to the electron angle distribution is completely dominated by the 
contribution of electron radiation, the other effects being almost flat and much smaller over the full 
range. However, if an acoplanarity cut is applied, the contributions due to muon radiation 
and up-down interference corrections become visible in the region of small electron scattering angles, where they 
amount to some per cent. Interestingly, the above contributions have the same sign in the 
$\mu^+ e^- \to \mu^+ e^-$ process (upper panel of the left plot) and sum up to contribute to 
the overall QED correction, whereas they tend to cancel in the $\mu^- e^- \to \mu^- e^-$
process (lower panel of the left plot). As far as the $d\sigma / d\theta_\mu$ cross section
is concerned, it can be noticed that, both in the absence and in the presence of 
acoplanarity cuts, all the contributions play a r\^ole, especially in the region of relatively large
muon angles around 4.5~mrad. In this region, the different sets of correction vary between 
5\% and 20\%, the dominant contribution being electron radiation, followed by 
the interference correction and muon radiation, respectively. Like for the
$d\sigma / d\theta_e$ cross section, 
all the corrections have the same sign in the final part of the spectrum 
of the  $\mu^+ e^- \to \mu^+ e^-$ process (upper panel of the right plot), whereas 
for the $\mu^- e^- \to \mu^- e^-$ process (lower panel of the right
plot) the positive up-down interference contributions 
tend to compensate the negative effect of the electron and muon radiation.
Similar considerations apply to the  $d\sigma / d t_{\mu\mu}$ and $d\sigma / d t_{ee}$ 
distributions shown in Fig.~\ref{Fig:Fig11}. All in all, these results indicate that 
all the sources of corrections have to be taken into account for precise predictions
for the differential cross sections.

\begin{figure}[ht]
\begin{center}
\includegraphics[width=0.5\textwidth]{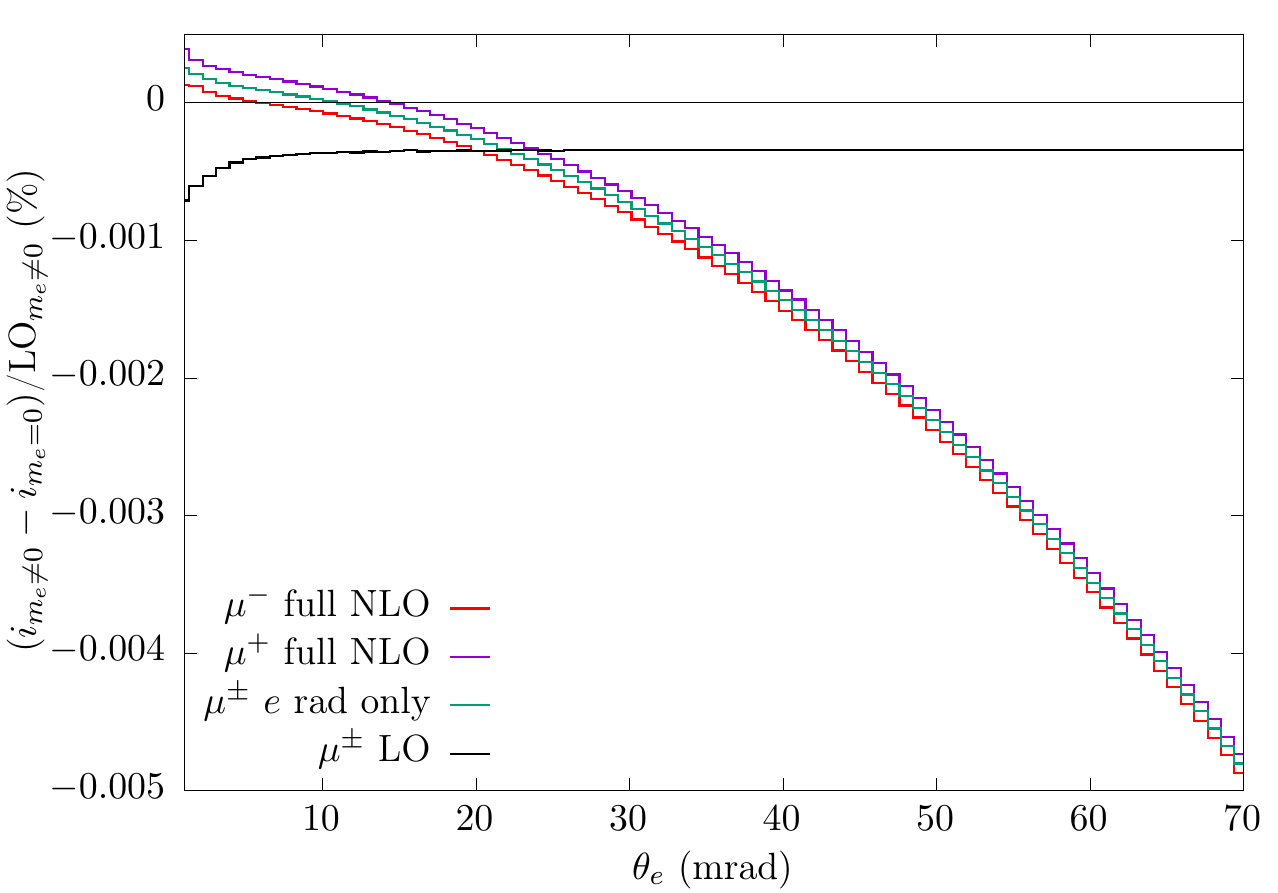}~~\includegraphics[width=0.5\textwidth]{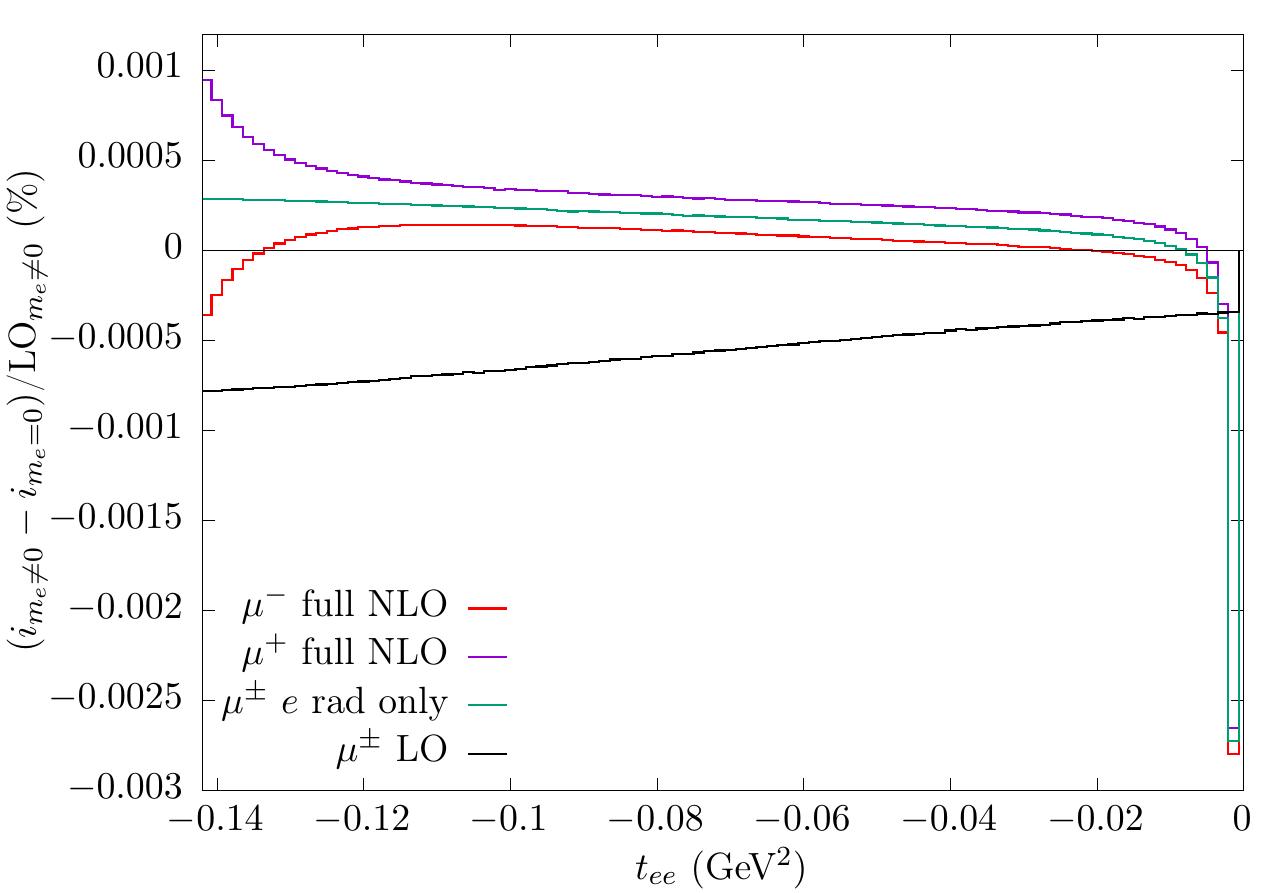}
\end{center}
\caption{\label{Fig:Fig12} Left plot: the contribution of 
the electron mass corrections to the cross section of the processes
$\mu^\pm e^- \to \mu^\pm e^-$, as a function of the electron scattering angle. Right plot: 
the same as in the left plot as a function of the squared momentum transfer $t_{ee}$. The predictions 
refer to Setup 1 defined in the text.}
\end{figure}

We now discuss finite electron mass corrections. We are interested in assessing
the (IR-safe) effects due to taking the limit of vanishing electron
mass in the virtual (and real soft-photon) amplitudes. In order to achieve the aim, we
adopt the following procedure to approximate the ingredients of the
complete NLO QED correction:

\begin{itemize}
\item we keep fully massive momenta. The choice is driven by the
  fact that with a massless electron its rest frame (i.e. the lab
  frame) could not be defined. We therefore keep fully massive phase-space
  and flux factor. We then choose to express the $2\to 2$ amplitudes
  in terms of the independent $s$ and $t$ Mandelstam invariants;

\item for virtual corrections, we completely reduce the tensor structures
  to scalar functions. We then retain exact $m_e$ dependency in
  those terms enhanced by logarithms of the artificial IR parameter
  $\lambda$. In the non-IR remainder, we express $u=2m_\mu^2-s-t$
  (i.e. discarding $2m_e^2$) and we
  neglect any $m_e^2$ contribution except in the arguments of the collinear logarithms;

\item for real soft-photon corrections, we proceed similarly: we
  keep exact $m_e$ dependency in terms proportional to
  $\log\left(\omega_s/\lambda\right)$ and neglect any $m_e$
  corrections in the remainder, except in the arguments of the
  collinear logarithms;

\item for real corrections, we do not make any approximation, because
  in this case finite mass corrections emerge from a delicate
  interplay between matrix elements and integration over the 3-body phase space.
\end{itemize}

We remark that with these choices our $m_e\to 0$ limit is always
independent of the IR parameters $\lambda$ and $\omega_s$ by construction and gives an
estimate of the effects due to neglecting $m_e$ in the one-loop
virtual amplitudes. We also stress that also in our massless limit terms proportional to
$\log^2\left(Q^2/m_e^2\right)$ are cancelled out between virtual and real soft-photon
corrections, as in the exact calculation.

The method outlined above has been applied in the computation of the 
full NLO QED corrections and of the gauge-invariant 
subset given by the electron-line {(virtual and real)} contributions.
The comparison between the exact results of  
Section~\ref{sec:qedexact} and those obtained in the $m_e\to 0$ limit
provides an IR-safe estimate of finite electron mass effects.

In Fig.~\ref{Fig:Fig12} we show our results for Setup 1, in per cent of the
fully massive LO differential cross section.
The predictions refer to both incoming $\mu^+$ and $\mu^-$ and are just shown 
for the electron scattering angle and the squared momentum transfer 
$t_{ee}$, for the sake of illustration. Similar results apply to the other distributions considered in our study. 

From Fig.~\ref{Fig:Fig12}, it can be noticed 
that the contribution of $m_e$-dependent terms to the LO cross section is almost flat 
and below the 10ppm level. In the considered Setup, the electron-mass corrections at  NLO contribute 
to $d\sigma / d\theta_e$ and $d\sigma / d t_{ee}$ 
in the range from a few to some $10^{-5}$, getting larger as
$\theta_e$ increases. We notice that the largest part
of the finite $m_e$ corrections is due to radiation from the electron
line only, the full correction lying around it. The extra corrections
with respect to the electron line only are dominated by up-down interference and
box diagrams. 
 
\begin{figure}[ht]
\begin{center}
\includegraphics[width=0.5\textwidth]{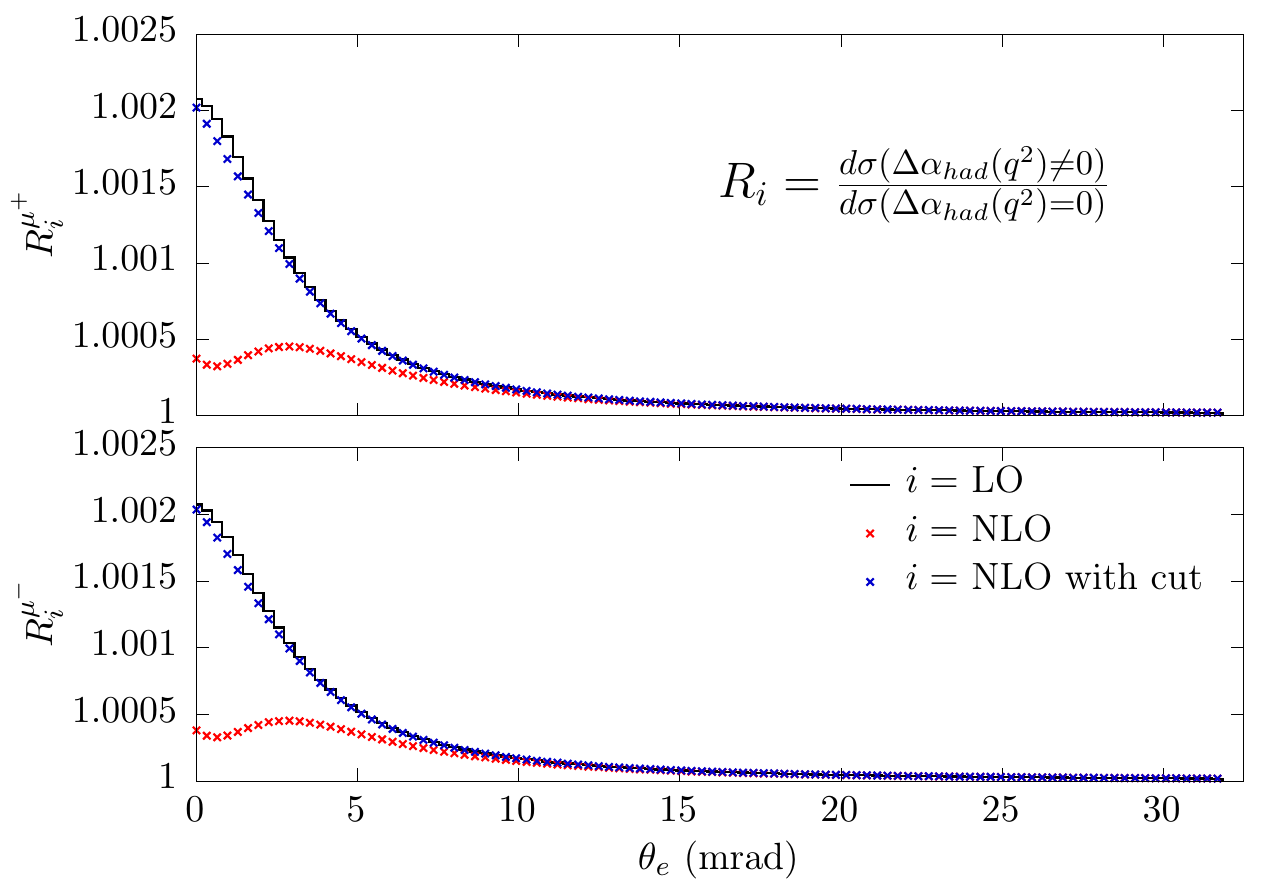}~~\includegraphics[width=0.5\textwidth]{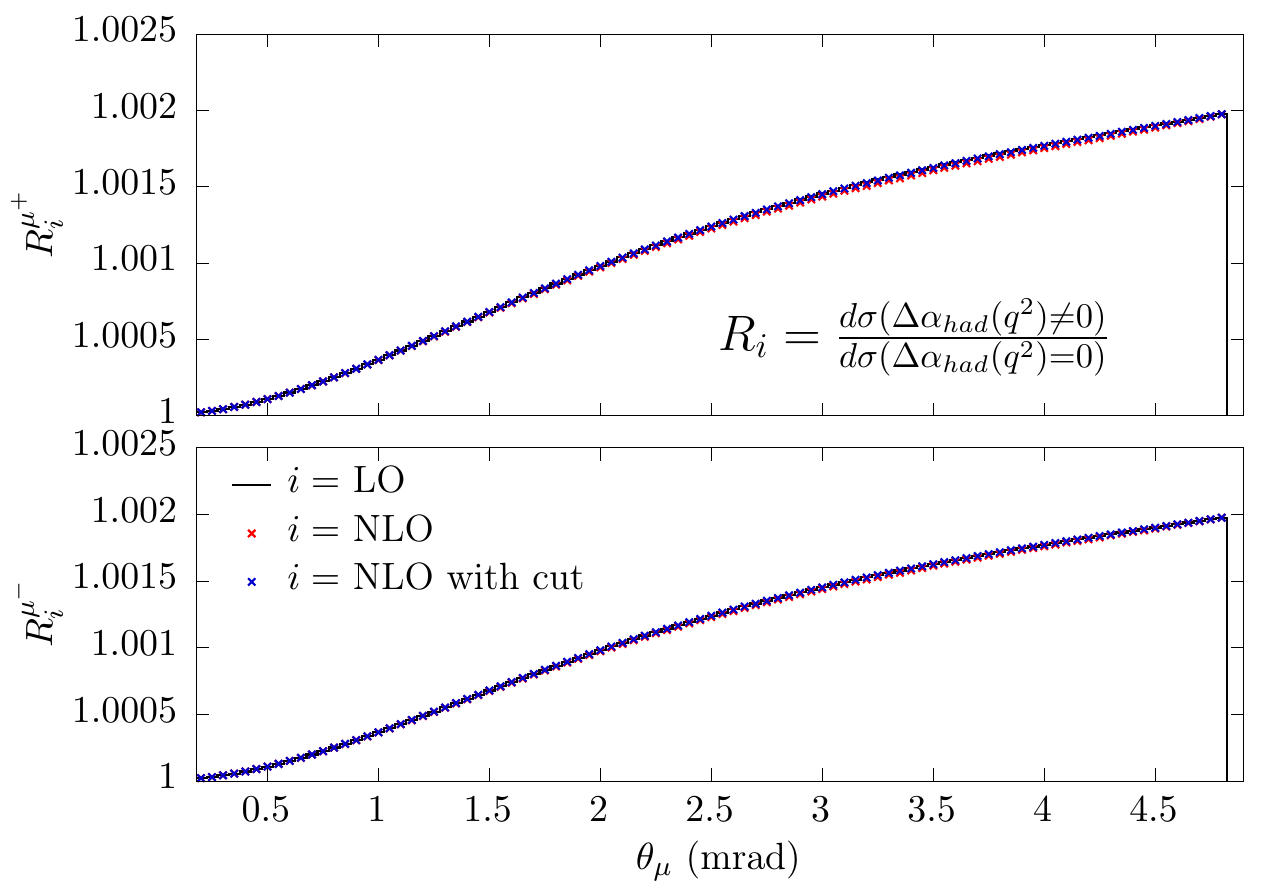}
\end{center}
\caption{\label{Fig:Fig13} Left plot: the ratios $R_i$ defined in the text for the process $\mu^+ e^- \to \mu^+ e^-$ 
(upper panel)  and the process $\mu^- e^- \to \mu^- e^-$ (lower panel), as a function of the electron scattering angle.
Right plot: 
the same as in the left plot as a function of the muon scattering angle. The results refer to 
Setup 2 and Setup 4 described in the text.}
\end{figure}

\begin{figure}[ht]
\begin{center}
\includegraphics[width=0.5\textwidth]{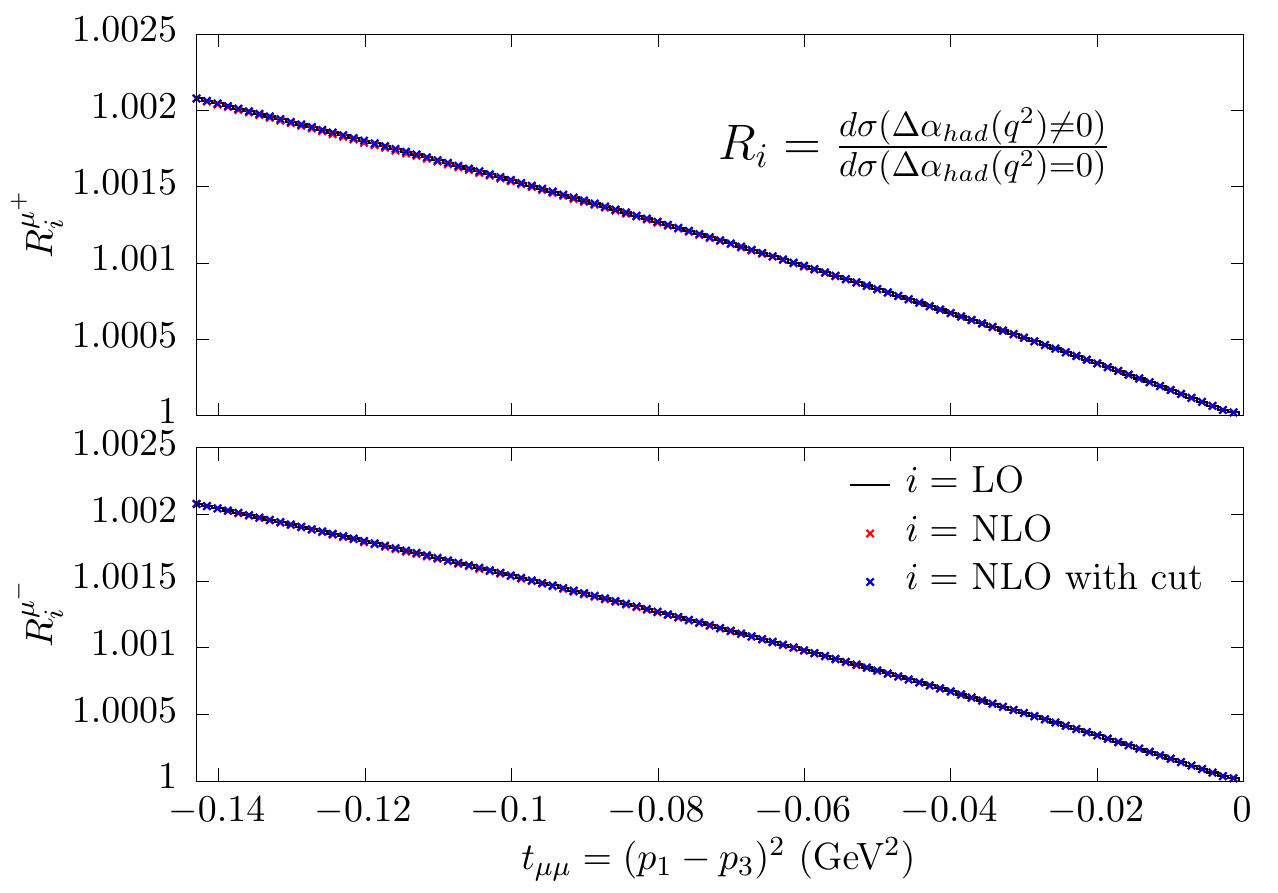}~~\includegraphics[width=0.5\textwidth]{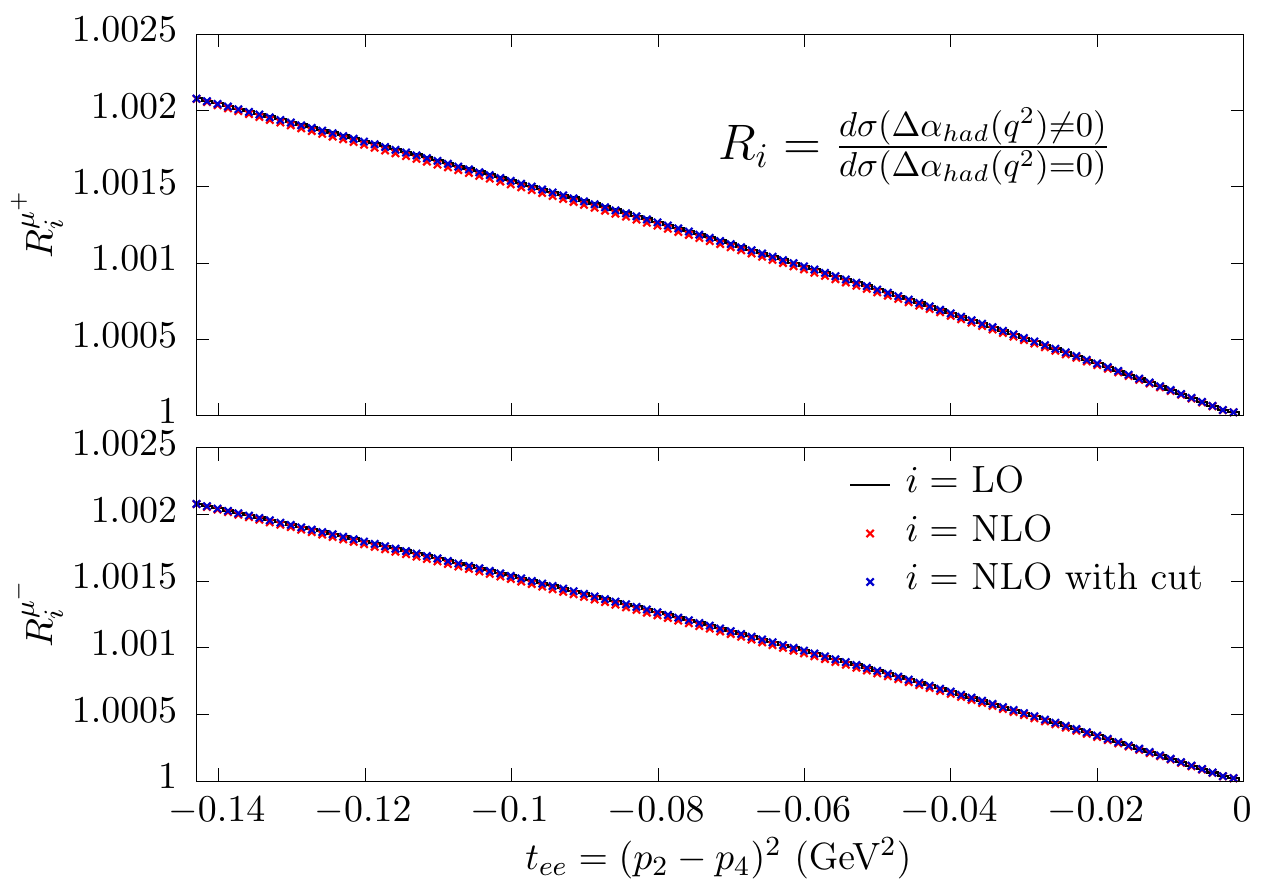}
\end{center}
\caption{\label{Fig:Fig14} The same as in Fig.~\ref{Fig:Fig13} as a function of the squared momentum transfer $t_{\mu\mu}$ (left plot) and $t_{ee}$ 
(right plot).}
\end{figure}

To conclude, we show in Fig.~\ref{Fig:Fig13} and in Fig.~\ref{Fig:Fig14} numerical results for the 
ratios defined as follows:
\begin{eqnarray} 
R_i \, = \, \frac{d\sigma_i (\Delta\alpha_\text{had.} (t) \neq 0) }{d\sigma_i (\Delta\alpha_\text{had.} (t) = 0)}
\label{eq:ratio}
\end{eqnarray}
According to Eq.~(\ref{eq:ratio}), $R_i$ is defined as the ratio between a generic cross section including
the contribution of the hadronic correction to the running of $\alpha$ and the same cross section 
without it. As such, this quantity gives the sensitivity of a given observable to the signal of interest 
for the MUonE experiment. We study how the ratios $R_i$ are affected by the QED corrections under 
the different selection criteria and for the two processes $\mu^\pm e^- \to \mu^\pm e^-$.

The results of this study are shown in Fig.~\ref{Fig:Fig13} as a function of the electron (left plot) 
and muon (right plot) scattering angles. Figure~\ref{Fig:Fig14} illustrates the situation as a 
function of the squared momentum transfer $t_{\mu\mu}$  (left plot) and $t_{ee}$ (right plot).  
As can be seen, there is not an appreciable difference between the results valid for the process 
initiated by positive muons (upper panels) and those involving negative muons (upper panels). 

From Fig.~\ref{Fig:Fig13}, it is particularly evident that, in the absence of elasticity conditions (Setup 2), the ratio as a function of the
 electron scattering angle is strongly affected by the contribution of QED radiation, especially in the region of 
 small angles, where the highest sensitivity, at the per mille level, is present. However, this sensitivity is largely recovered by applying 
 an acoplanarity cut (Setup 4), which removes the contribution of the radiative processes. On the other hand, the ratio 
 as a function of the muon scattering angle appears to be particularly robust under the contribution of the
 radiative corrections and applied cuts. The same property is shared by the ratios as functions of
 $t_{\mu\mu}$ and $t_{ee}$, as can be seen from Fig.~\ref{Fig:Fig14}.

\section{Conclusions}
\label{sec:conclusions}

In this work, we have computed the full set of NLO corrections to $\mu^\pm e^- \to \mu^\pm e^-$ scattering in the SM. The study is motivated 
by the recent proposal of measuring the effective electromagnetic coupling constant 
in the space-like region by using this process (MUonE experiment).
From the measurement of the 
hadronic contribution to the running of $\alpha_\text{QED}$, the leading hadronic contribution to the muon anomaly 
can be derived according to an alternative approach to 
the standard time-like evaluation of $a_\mu^\text{HLO}$.

In view of the challenging target accuracy (at the 10ppm level),
we have computed the NLO QED and purely weak corrections 
without any approximation and developed a 
corresponding fully differential MC code, which is available for first experimental studies.

We have performed a comprehensive phenomenological analysis of the $\mu e \to \mu e$ process at NLO accuracy 
under different event selection conditions, in order to study the dependence of the radiative corrections on the applied cuts. 
To this end, we have provided detailed results both at the level of integrated cross sections and distributions.

We have shown that the NLO electroweak corrections are largely dominated by the QED effects, the contribution of the
NLO purely weak corrections being typically well below $10^{-5}$ and hence negligible. Only the tree-level 
$\gamma+Z$ contribution plays a r\^ole, as its contribution can reach the $10^{-5}$ level in particularly important 
kinematical regions. Therefore, a first conclusion of our study is that at least the LO electroweak contributions 
are necessary to achieve the required theoretical accuracy.

We have also scrutinised the accuracy in the calculation of the leptonic corrections to the vacuum polarisation,
since, according to the proposed strategy of the MUonE experiment, the leading hadronic contribution to the muon 
anomaly is obtained after subtraction of the leptonic contribution to the running of the QED coupling. 
We have shown that the leptonic corrections beyond the one-loop approximation, i.e at two-loop and three-loop
accuracy, are of the order of $10^{-5}$ and $10^{-7}$, respectively, and therefore are not a limitation to 
the proposed experimental procedure.

We have examined the impact of the photonic corrections to all the observables of experimental interest, 
by considering some variants of possible experimental cuts. We have pointed out that, by assuming the 
presence of acceptance cuts only, the integrated cross sections receive a moderate correction, of about 
5\%. By imposing an elasticity condition given by a tight acoplanarity cut, the QED corrections to the 
integrated cross sections change sign and rise to the 10\% level, as a consequence of the increased
importance of soft photon emission.

We have analysed how the differential cross sections are affected by the QED corrections under the different selection criteria.
 We have shown that the electron angle distribution receives an
 extremely important and strongly varying 
 correction,  when no elasticity cuts are applied.  This behaviour is due to a large kinematical effect induced 
 on the electron scattering angle by the processes accompanied by real photon emission. However, 
 in such an experimental configuration, the corrections to the other distributions are much less pronounced 
 and slowly varying. We have demonstrated that a simple acoplanarity cut is of great help in getting 
 rid of most of the radiative processes, which populate the region of relatively small muon angles, 
 and singles out elastic events.
 The application of this cut tends to stabilise the QED contribution to the electron angle distribution, 
 giving rise to a flat and less significant correction of the order of 10\%. However, as a drawback, 
 the acoplanarity cut enhances the contribution of the photonic corrections to the other differential 
 cross sections, because of the growing importance of soft photon radiation. More sophisticated elasticity
 conditions are being studied at the experimental level and simulated
 accordingly, in order to optimise
 the dependence of the radiative corrections on the applied cuts.
 
 We have studied how the different gauge-invariant subsets of photonic corrections contribute to the
 overall QED contribution. We have shown that, in general, the radiation by the electron leg is the 
 dominant contribution but also emphasised that all the sources of corrections play a r\^ole 
 for a correct description of the distributions. In particular, we
 have shown that the corrections 
 due to up-down interference and box contributions are of opposite sign for the processes
 involving opposite charge muons, thus explaining the difference between the corrections
 affecting the  $\mu^+ e^- \to \mu^- e^-$ and  $\mu^- e^- \to \mu^- e^-$ scattering.  
 In particular, we have pointed out that the interference corrections provide a significant 
 contribution for small electron scattering angles or, equivalently, at relatively large 
values of the squared momentum transfer, which defines the signal region of the MUonE experiment.
 
We have also given an estimate of the finite electron-mass corrections, to show that they 
are not negligible at NLO accuracy and lie in the range from a few
to some $10^{-5}$, being dominated by the corrections from the electron leg.

In conclusion, we have also pointed out that, independently of the
applied cuts, the cross section as a function of the muon scattering angle is a particularly robust observable 
under the contribution of the radiative corrections for the extraction of the signal of interest. However, also
the electron angle distribution can be used to this end, if appropriate elasticity cuts are applied.

The results here presented represent the first step towards the realisation of a high-precision theoretical tool necessary for the 
analysis of the data of the MUonE experiment. In the near future, we plan to match the 
NLO QED corrections to $\mu e \to \mu e$ with the contribution due to multiple photon emission, 
following the formulation successfully applied to Bhabha scattering and QED processes 
at flavour factories~\cite{Balossini:2006wc,Balossini:2008xr,Actis:2010gg}, 
Drell-Yan processes at hadron colliders~\cite{CarloniCalame:2006zq,CarloniCalame:2007cd} and Higgs 
boson decay into four leptons~\cite{Boselli:2015aha}. Over the longer term, we are interested in developing a MC
event generator including NNLO corrections and resummation, which will be ultimately needed for data analysis.

\acknowledgments
We are sincerely grateful to all our MUonE colleagues for stimulating collaboration and 
many useful discussions, which are the framework of the present study. 
We are indebted to Matteo Fael and Massimo Passera for providing us
with the results of their calculation used in the tuned comparison
shown in the paper.

We thank for its hospitality and its partial support the Mainz Institute for Theoretical Physics
(MITP), where part of the present work has been carried out
during the topical workshop {\it ``The evaluation of the
leading hadronic contribution to the muon anomalous magnetic moment''}
held in February 2018.

\bibliographystyle{JHEP}
\bibliography{muone-JHEP}

\end{document}